\documentclass[prb,twocolumn,showpacs,preprintnumbers,amsmath,amssymb]{revtex4} 
\usepackage{graphicx}
\usepackage{dcolumn}
\usepackage{epsfig,graphics,amsbsy,calc,latexsym}
\usepackage[italian,english]{babel}
\usepackage{amsmath, amssymb}
\usepackage{bm}

\def\tc{$T_c$\ }

\def\ra{\rightarrow}
\def\ds{\displaystyle}

\def\sss{\scriptscriptstyle}
\def\ts{\textstyle}
\def\nn{\nonumber}

\def\sw{$s$--wave\ }

\def\kk{\mathbf{k}}

\def\RR{\mathbf{R}}
\def\UU{{\bf U}}
\def\TT{{\bf T}}

\def\ee{\bm{\epsilon}}
\def\eps{\varepsilon}
\def\um{{\ts \frac{1}{2}}}
\def\de{{\partial}}
\def\detau{{\partial_\tau}}
\def\RRp{{\mathbf{R}^{(p)}}}
\def\RRh{{\mathbf{R}^{(h)}}}
\def\btc{\begin{tabular}{l} }
\def\etc{\end{tabular} }

\def\ord{\mathcal{O}}

\def\U{\mathcal{U}}
\def\J{\mathcal{J}}

\def\A{\mathcal{A}}
\def\D{\mathcal{D}}
\def\Z{\mathcal{Z}}
\def\vac{|\mathrm{vac}\rangle}

\def\ph#1{{\phantom{#1}}}
\def\eq#1{(\ref{#1})}
\def\ceq#1{Eq.~(\ref{#1})}
\def\eqs#1#2{(\ref{#1}-\ref{#2})}
\def\ceqs#1#2{Eqs.~(\ref{#1}-\ref{#2})}
\def\sec#1{Sec.~\ref{#1}}
\def\fig#1{Fig.~\ref{#1}}
\def\ident{\mathbf{1}}

\begin{document}
\title{Rotationally-invariant slave-bosons for Strongly Correlated Superconductors}
\author{A. Isidori$^{1}$ and M. Capone$^{2,1}$}

\affiliation{$^{1}$ Dipartimento di Fisica, Universit\`a di Roma ``La Sapienza'', Piazzale A. Moro 2, 00185, Rome, Italy} 
\affiliation{$^{2}$ SMC, CNR-INFM and ISC-CNR, Piazzale A. Moro 2, 00185, Rome, Italy} 

\begin{abstract} 
We extend the rotationally invariant formulation of the slave-boson method to superconducting states. This generalization,
building on the recent work by Lechermann et al. [Phys. Rev. B {\bf 76}, 155102 (2007)], allows to study superconductivity in strongly correlated systems. We apply the formalism to a specific case of strongly correlated superconductivity, as that found in
a multi-orbital Hubbard model for alkali-doped fullerides, where the superconducting pairing has phonic origin, yet it has been shown to be favored by strong correlation owing to the symmetry of the interaction. The method allows to treat on the same footing the strong correlation effects and the interorbital interactions driving superconductivity, and to capture the physics of strongly correlated superconductivity, in which the proximity to a Mott transition favors the superconducting phenomenon.

\end{abstract}
 
\date{\today} 
\maketitle 

\section{Introduction}\label{sec:introduction}

The theoretical description of strongly correlated systems and of the prototypical models introduced to understand their behavior plays a central role in  
modern many-body theory. Even if the number of materials of interest in which the mutual interaction 
 between electrons has been identified as relevant is now countless,
there is no doubt that the main trigger for the development 
of the correlated-electron field has been the discovery of high-temperature superconductivity 
in doped correlated insulators such as the copper oxides.
Yet, the link between strong correlation and high-temperature superconductivity has
not been established unambiguously, which prompts for theoretical methods able to
describe the superconducting phenomenon in the presence of strong electron correlations.

One of the main reasons why strongly correlated systems and their properties are,
 at the same time, interesting and hard to solve is that they are intrinsically out 
 of weak-coupling regimes, where a perturbative expansion can be performed.
Starting from an uncorrelated system and imagining to continuously increase the 
degree of correlation, the relevance of local repulsion gradually introduces constraints 
to the electronic motion, leading eventually to the localization of the
 carriers (Mott transition). Thus a proper method for correlated electrons should be able 
 to introduce local constraints onto an otherwise uncorrelated state, that would be naturally 
 delocalized, i.e., spatially unconstrained.

A popular strategy which formally implements this point of view is based on slave bosons\cite{general_slaveboson,Kotliar_Ruckenstein:86}. Within these approaches, the Hilbert space is enlarged to include, besides fermionic degrees of freedom associated to Landau quasiparticles, suitable extra degrees
of freedom of bosonic character which are typically related to local states. The auxiliary (slave) degrees of freedom are then
treated in a mean-field approximation, leading to an effective low-energy theory for the quasiparticles.
The high-energy physics can only be recovered introducing fluctuations of the fields describing the auxiliary particles\cite{castellaniraimondi}.
We can see this strategy as a way to enforce a local point of view, which is expected to be 
correct in very strong coupling, starting from delocalized non-interacting states.
In its most popular version, introduced by Kotliar and Ruckenstein\cite{Kotliar_Ruckenstein:86}, 
one defines one boson for each local configuration (namely, $|0\rangle$, $\left|\uparrow \right\rangle$, $\left|\downarrow \right\rangle$, $\left|\uparrow \downarrow \right\rangle$), and the equivalence between 
the physical Hilbert space and the new extended space is enforced by imposing constraints 
which, as we shall discuss below, imply that the local configurations 
should be coherently labeled by the fermionic and bosonic degrees of freedom, and that 
precisely one boson should be present on each lattice site.

Yet, as thoroughly discussed in Ref.~\onlinecite{Lechermann_Georges:07}, the standard 4-boson 
representation\cite{Kotliar_Ruckenstein:86} for the single-band Hubbard model,
as well as its simplest generalizations to multi-orbital models\cite{Fresard_Kotliar:97}, are not suitable to handle
arbitrary forms of the interaction Hamiltonian, characterized by terms which cannot be put in the form of density-density interactions, such as exchange interactions associated to the Hund's rule coupling.
Furthermore, even for pure density-density interactions, the Kotliar-Ruckenstein approach still remains inadequate to handle charge symmetry breaking order parameters such as the superconducting one in the Hubbard model. 
Slave-boson approaches to superconductivity in the Hubbard model have indeed mostly used the approximately equivalent strong-coupling t-J model, and have been based on specific assumptions\cite{kopp}.

The first attempt in overcoming the inadequacy of Kotliar-Ruckenstein's representation was made by 
Li \textit{et al.}\cite{Li_Wolfle:89}, who proposed a spin-rotation invariant slave-boson formulation
of the single band Hubbard model, while in Ref.~\onlinecite{Fresard-Wolfle:92} Fr\'esard and W\"olfle introduced a more general 
representation for single-band models, in which spin and charge degrees of freedom are treated on the same footing and 
rotational invariance involves both spin and particle-hole transformations. Although the formalism presented by these authors 
refers only to a 4-state system, with the slave-boson fields labeled in correspondence with the specific $SU(2)\otimes SU(2)$ generators of spin
and particle-hole rotations, it already has all the ingredients required for describing systems with local superconducting pairing.
Such method has been indeed applied to the single-band attractive Hubbard model in Ref.~\onlinecite{Bulka:96}.
Developing the ideas of these pioneering works in a more systematic way, Lechermann \textit{et al.}\cite{Lechermann_Georges:07} finally built a completely basis-independent slave-boson formalism, suitable to describe, within a generic multi-orbital model, 
any arbitrary form of local interaction. As it is found for Kotliar-Ruckenstein's approach, it is worth mentioning that, at mean-field level,
such formalism turns out to be equivalent\cite{Bunemann:07} to analogous extensions of the Gutzwiller approach\cite{Bunemann:98,Fabrizio}.

While the possibility of extending the formalism to superconducting states is mentioned in Ref.~\onlinecite{Lechermann_Georges:07}, the explicit derivation is limited to normal solutions, imposing no charge symmetry breaking. 
In the present work, instead, we lift this restriction and consider explicitly the more general 
case of full rotational-invariance under any local transformation of the electronic degrees of freedom, and we apply the formalism to solve a three-orbital model which has been proposed to describe alkali-doped fullerides\cite{Capone-Science,CaponeRMP}. Besides its relevance to the fullerides, the model has important properties that led us to choose it as an optimal benchmark for our method.
The model has indeed been shown to present ``strongly correlated superconductivity''\cite{Capone-Science}, i.e., the enhancement of phonon mediated superconductivity in the proximity of a Mott transition. The key of the phenomenon is that a small attraction which involves orbital and spin degrees of freedom is not screened when charge fluctuations are frozen by strong correlations. This leads to an enhancement of superconductivity, since the unscreened attraction now acts on strongly renormalized quasiparticles with a larger effective density of states. 
This effect has been identified using Dynamical Mean-Field Theory (DMFT)\cite{DMFT}, which fully takes into account local quantum fluctuations, but it has not been reproduced by ordinary slave-boson methods due to the difficulties in treating interactions which are not of the charge-charge form, such as those driving superconductivity in the model we are dealing with.
In this light, the model is an ideal test ground of the ability of the rotationally-invariant slave boson method in accurately treating general forms of interactions. On the other hand, the model has only local (on-site) interactions, which simplifies the approach.

The paper is organized as follows. In Sec.~\ref{sec:formalism} we introduce the rotationally-invariant slave boson method 
for models with local superconducting pairing.
In Sec.~\ref{sec:fullerides} we present the multi-orbital model used for the description of fullerenes, illustrating
the way it can be solved by means of the slave boson approach. 
In Sec.~\ref{sec:results} we present the results obtained with our method, and finally 
Sec.~\ref{sec:conclusions} is dedicated to concluding remarks and perspectives.

\section{The general formalism}\label{sec:formalism}

In this section we will explicitly extend the rotationally-invariant slave-boson
formalism introduced by Lechermann \textit{et al.}\cite{Lechermann_Georges:07} to the possibility of describing superconducting states. To facilitate the reading and the comparison with the formalism of Ref.~\onlinecite{Lechermann_Georges:07}, 
whenever possible we shall use the same notation for corresponding quantities.

\subsection{Motivations}

Without entering in details, we shall first provide a brief reminder on slave-boson formulations, 
in order to face the difficulties encountered with non-invariant approaches such as Kotliar-Ruckenstein's one.

In a generic multi-orbital model the local Hilbert space of electronic states is defined as 
the set of all the possible ``atomic'' configurations at a given lattice site (for simplicity,
we will drop site indices throughout this section); a natural choice for the basis set of
this space is provided by the $2^M$ Fock states 
\begin{equation}
 |n\rangle \equiv  
 \left(d_1^\dagger\right)^{n_1} \cdots \left(d_M^\dagger\right)^{n_M} \vac, \qquad [n_\alpha = 0,1]
\end{equation}
where $\alpha=1,\ldots,M$ are the local orbital species and $d^\dagger_\alpha$
the corresponding electron-creation operators. A slave-boson representation is then constructed
by mapping the local Hilbert space $\mathcal{H}$ (e.g., the Fock states $|n\rangle$) onto an ``enlarged'' Hilbert space 
$\underline{\mathcal{H}}$ generated by the tensor products of boson operators $\phi_\mu^\dagger$ and auxiliary fermion operators $f^\dagger_\alpha$:
\begin{equation}
\mathcal{H} : \{ |n\rangle \} \; \longmapsto \; \underline{\mathcal{H}} : \left\{\ts \prod_\mu \left(\phi_\mu^\dagger\right)^{N_\mu} \!\vac 
  \otimes |n\rangle_f \right\}. \nn
\end{equation}
In the above expression $|n\rangle_f$ refers to the Fock states generated by the auxiliary fermions $f^\dagger_\alpha$,
which correspond to quasiparticle (QP) degrees of freedom: their presence ensures the possibility of describing 
Fermi liquid properties within the auxiliary-fields representation (note that the orbital basis 
for quasiparticle degrees of freedom may not coincide, in general, with that of the physical electron 
operator $d^\dagger_\alpha$). On the other hand, an arbitrary number 
of auxiliary bosons $\phi_\mu$, for each species $\mu$, can in principle be present in the enlarged Hilbert space, unless some
constraints, which characterize the specific form of the slave-boson representation, are imposed to its states.
In other words, a given representation is defined by the way in which the auxiliary states are selected
out of the enlarged space $\underline{\mathcal{H}}$ in order to represent uniquely the original physical states $|n\rangle$,
\begin{equation}
\begin{array}{c}
 |n\rangle \; \longmapsto \; |\underline{n}\rangle, \\[.3cm]
\ds |\underline{n}\rangle \equiv  \!\!\sum_{\{N_\mu\},\,m}\!\! K(n, \{N_\mu\}, m) \,
  \prod_\mu \left(\phi_\mu^\dagger\right)^{N_\mu} \!\vac  \otimes |m\rangle_f.
\end{array} 
\end{equation} 
Needless to say, the choice of a specific representation $K$, apart from being consistent with the above uniqueness assumption,
must also provide some simplifications in the (local) interaction Hamiltonian: the whole purpose of
introducing auxiliary bosons is indeed the possibility of writing local interactions as a sum of quadratic terms in the boson fields,
at the expense of a larger number of degrees of freedom and a more complex structure of hopping terms.

In multi-orbital generalization of Kotliar-Ruckenstein's approach, 
$2^M$ boson fields $\phi_\mu \equiv \phi_n$ are introduced in correspondence with the original Fock states $|n\rangle$,
and the representation of such states in the enlarged Hilbert space reads
\begin{equation}
|\underline{n}\rangle \equiv \phi_n^\dagger \vac \otimes |n\rangle_f. \label{eq:density_phys_state}
\end{equation}
The ``physical states'' in $\underline{\mathcal{H}}$ are therefore those states containing exactly one boson and
whose quasiparticle content $|n\rangle_f$ matches the Fock configuration associated to the boson field $\phi_n$.
From \ceq{eq:density_phys_state} it should be evident the non-invariant nature of this representation
under rotations of the quantization basis. Consider, in fact, an $SU(M)$ rotation of the orbital indices\cite{note3}
\begin{equation}
 d^\dagger_\alpha = {\ts \sum_\beta} U_{\alpha\beta} \tilde{d}^\dagger_\beta,\qquad 
 f^\dagger_\alpha = {\ts \sum_\beta} U_{\alpha\beta} \tilde{f}^\dagger_\beta.
\end{equation} 
This rotation will induce a corresponding unitary transformation on both physical and QP Fock states,
$|n\rangle = \sum_m \mathcal{U}(U)_{nm} |\widetilde{m}\rangle$, so that the representation of physical states 
would now read
\begin{eqnarray}
|\underline{\widetilde{m}} \rangle &=& \sum_{n\,m'} \left( \mathcal{U}^\dagger_{mn} \phi_n^\dagger \mathcal{U}_{nm'} \right) \vac 
 \otimes |\widetilde{m}'\rangle_{\tilde{f}} \nn \\ 
 &=& \sum_{m'} \tilde{\phi}^\dagger_{mm'} \vac \otimes |\widetilde{m}'\rangle_{\tilde{f}}.
\end{eqnarray}
In the new orbital basis, therefore, the slave-boson representation do not retain
its original form, and more specifically the definite relation between physical states
and their quasiparticle content no longer holds. As discussed in Ref.~\onlinecite{Fresard-Wolfle:92}, indeed,
only disentangling physical and quasiparticle degrees of freedom it becomes possible to formulate rotationally-invariant
slave-boson representations. 

As a consequence of its basis-dependent nature, Kotliar-Ruckenstein's approach can be 
applied only to systems whose local Hamiltonian can be written, in an appropriate basis, in terms of
purely orbital-density operators 
$\hat{n}_\alpha = d^\dagger_\alpha d_\alpha$, 
\begin{equation}
H_\mathrm{loc} = \sum_\alpha \epsilon^0_\alpha \hat{n}_\alpha + \sum_{\alpha\beta} W_{\alpha\beta}\, \hat{n}_\alpha \hat{n}_\beta,
\end{equation}
i.e., when the Fock states $|n\rangle$ are eigenstates of $H_\mathrm{loc}$. In this case, 
the representation of $H_\mathrm{loc}$ in the enlarged Hilbert space can be easily written as
a free-boson Hamiltonian, 
\begin{equation}
\underline{H}_\mathrm{loc} = \sum_n E_n \phi^\dagger_n \phi_n, 
\end{equation}
with 
$E_n = \sum_\alpha \epsilon^0_\alpha n_\alpha + \sum_{\alpha\beta} W_{\alpha\beta}\, n_\alpha n_\beta$.
We remark, however, that the definite relation imposed between quasiparticle degrees of freedom 
and the (physical) Fock content of boson fields  inhibits the development of those spontaneous symmetry-breaking order-parameters
that cannot be expressed in terms of orbital-density operators (e.g., superconductivity, 
magnetization perpendicular to the spin-quantization axis, etc.).

\subsection{Representation of physical states}

The electron Hamiltonian for a generic multi-orbital model with purely local interactions is given by
\begin{eqnarray}
 H &=& H_\mathrm{kin} + \sum_i H_\mathrm{loc}[i], \label{eq:generic_Ham}\\
 H_\mathrm{kin} &=& \sum_\kk \sum_{\alpha\beta}
  \epsilon_{\alpha\beta}(\kk)d_{\kk\alpha}^\dagger d_{\kk\beta}, \label{eq:H_kin_gen}
\end{eqnarray}
where  all the local terms, including the chemical potential and
the orbital energy levels, are included in $H_\mathrm{loc}$, so
that $\sum_\kk \epsilon_{\alpha\beta}(\kk)=0$.

In comparison to the previous subsection, we choose here,
as the basis set for the (physical) local Hilbert space,  a generic set
of states $\{|A\rangle\}$ (not necessarily Fock states) that are eigenstates of the local
particle-number operator $\hat{n}^{(d)} = \sum_{\alpha=1}^M d_{\alpha}^\dagger d_{\alpha}$, with eigenvalues $N_A$.
Eventually, among these sets, we can choose the eigenstates
$\{|\Gamma\rangle\}$ of the local Hamiltonian, since $H_\mathrm{loc}$
commutes with the local number operator. The basis set for quasiparticle states, instead, 
is still given by the Fock states $|n\rangle_f$ generated by the auxiliary fermion
operators $f^\dagger_\alpha$.

As discussed previously in pointing out the limitations
of Kotliar-Rucken\-stein's approach, the key ingredient in constructing 
rotationally-invariant slave-boson representations is to disentangle physical and quasiparticle 
degrees of freedom\cite{Lechermann_Georges:07}.
Therefore, we introduce a set of auxiliary boson fields $\phi_\mu \equiv \phi_{An}$
associated, in principle, to each pair $( |A\rangle,\, |n\rangle_f )$ of physical and quasiparticle states,
without assuming any a priori relation between those states in the enlarged Hilbert space representation. 
Depending on the phases one takes into account, however,
there exist some limitations in the possible $\underline{\mathcal{H}}$--states $\phi_{An}^\dagger \vac \otimes |n\rangle_f$
which can figure in the representation of a physical state $|A\rangle$,
\begin{equation}
 |\underline{A}\rangle \,\propto\, 
 \sum_n \phi_{An}^\dagger \vac \otimes |n\rangle_f.
\end{equation}
Indeed, if we limit to
normal phases as in Ref.~\onlinecite{Lechermann_Georges:07}, it is sufficient to consider, for a given state $|A\rangle$, only 
those states $|n\rangle_f$ which have exactly the same number of particles of $|A\rangle$; in other words,
physical states with a definite number of electrons are represented by a superposition of auxiliary states characterized
by the same number of quasiparticles. On the other hand, when allowing for the spontaneous breaking of particle-number conservation, 
as in superconducting states, we need to consider, for each state $|A\rangle$, all the Fock states $|n\rangle_f$ 
characterized by $$\left[ \sum_{\alpha=1}^M n_\alpha - N_A \right] (\mathrm{mod\,} 2) =0,$$ i.e.,
all the $2^{M-1}$ quasiparticle states with the same statistics of $|A\rangle$. 
While the former representation is invariant only under rotations of the QP basis
that are block-diagonal in the quasiparticle occupation number $\sum_{\alpha=1}^M n_\alpha$,
the latter is invariant under a larger class of QP
rotations, represented by all the unitary transformations that preserve the statistics of quasiparticle states: 
in such representation, the quasiparticle 
number operator is no longer a conserved quantity, and its expectation value  
does not correspond to any physical observable, as particle-hole transformations may change its value.

The explicit representation of $|A\rangle$ 
will then read, in the fully-invariant formalism,
\begin{equation}
 |\underline{A}\rangle \equiv \frac{1}{\sqrt{2^{M-1}}}
 \sum_n \phi_{An}^\dagger \vac \otimes |n\rangle_f, \label{eq:phys_ket}
\end{equation}
with the sum running over the $2^{M-1}$ QP Fock states whose particle-number parity equals that of $|A\rangle$.
It is worthwhile to remark that the above representation cannot be further enlarged, including, for example,
in the definition of $|\underline{A}\rangle$, the remaining quasiparticle states with opposite statistics.
This, in fact, would lead to unphysical results such as non-vanishing expectation values of 
odd numbers of fermion operators.

\subsubsection*{Constraints}

In order to characterize uniquely the physical states among all
the states of the enlarged Hilbert space $\underline{\mathcal{H}}$, it is necessary and
sufficient that the selected states satisfy, as operator
identities, the following constraints:
\begin{eqnarray}
 \sum_{An} \phi_{An}^\dagger \phi_{An} &=& 1, \label{eq:const1} \\
 \sum_{A} \sum_{nn'} \phi_{An}^\dagger \phi_{An'}
   \langle n'| f_\alpha^\dagger f_{\alpha'} |n \rangle &=& f_\alpha^\dagger f_{\alpha'}, \label{eq:const2}\\
 \sum_{A} \sum_{nn'} \phi_{An}^\dagger \phi_{An'}
   \langle n'| f_\alpha^\dagger f_{\alpha'}^\dagger |n \rangle
   &=& f_\alpha^\dagger f_{\alpha'}^\dagger. \label{eq:const3}
\end{eqnarray}
The first two types of constraints are already present in the normal-phase
formalism of Lechermann \textit{et al.}\cite{Lechermann_Georges:07}: the first equation, indeed, limits
the physical subspace of $\underline{\mathcal{H}}$ to one-boson states only, while the second
set of constraints ensures the rotational invariance of \eq{eq:phys_ket} under QP rotations that preserve quasiparticle number. 
On the other hand, the last set of constraints promotes the rotational invariance to particle-hole rotations,
enabling the non-conservation of quasiparticle number. 

It is worthwhile to remark that, for single-band models ($M=2$), the above equations reduce to the same
set of constraints characterizing the $SU(2)\otimes SU(2)$ spin-charge 
invariant formalism introduced in Refs.~\onlinecite{Fresard-Wolfle:92} and \onlinecite{Bulka:96}, 
as long as the appropriate changes in notation are made. For this purpose, Ref.~\onlinecite{Bulka:96}
provides a useful link between our notation and the one presented in Ref.~\onlinecite{Fresard-Wolfle:92}.

A more compact form of \ceqs{eq:const1}{eq:const3} will be derived in \sec{sec:gauge}, where we shall relate the role of constraints to
the gauge group structure of the slave-boson representation.

\subsection{Physical electron operator}

The representation of the physical electron creation operator in
the enlarged Hilbert space is defined by
\begin{equation}
 \underline{d}_\alpha^\dagger |\underline{B} \rangle =
 \sum_A \langle A| d_\alpha^\dagger |B\rangle\, |\underline{A} \rangle.
\end{equation}

When the constraints \eqs{eq:const1}{eq:const3} are satisfied exactly, its expression in
terms of bosons and quasiparticle operators reads
\begin{eqnarray}
 \underline{d}_\alpha^\dagger &=& \frac{1}{M} \sum_{AB,\,nm,\,\beta}
 \langle A| d_\alpha^\dagger |B\rangle\,  \phi_{An}^\dagger \phi_{Bm} \; \times \nn \\
 & & \times \; \left[\langle n| f_\beta^\dagger |m \rangle f_\beta^\dagger +
 \langle n| f_\beta |m \rangle f_\beta \right],
\end{eqnarray}
with the normalization factor $1/M$ coming from the following relation:
\begin{equation}
 \sum_{p,\,\beta} \left[\langle n| f_\beta^\dagger |p \rangle f_\beta^\dagger |p \rangle +
 \langle n| f_\beta |p \rangle f_\beta |p \rangle \right] = M |n \rangle.
\end{equation}
We can thus summarize the non-diagonal relation between physical and quasiparticle degrees
of freedom in the form
\begin{equation}
 \underline{d}_\alpha^\dagger \,=\, R^{(p)}_{\alpha\beta}[\phi]^* f_\beta^\dagger 
 \,+\, R^{(h)}_{\alpha\beta}[\phi] f_\beta \label{eq:R-matrices}
\end{equation}
(summation over repeated indices is implied), where we have defined the $R$-matrix operators as 
\begin{eqnarray}
 R^{(p)}_{\alpha\beta}[\phi]^* &=&  \frac{1}{M} \sum_{AB,\,nm}
 \langle A| d_\alpha^\dagger |B\rangle\, \phi_{An}^\dagger \phi_{Bm} \langle m| f_\beta |n \rangle, \qquad \\
 R^{(h)}_{\alpha\beta}[\phi] &=&  \frac{1}{M} \sum_{AB,\,nm}
 \langle A| d_\alpha^\dagger |B\rangle\, \phi_{An}^\dagger \phi_{Bm} \langle m| f_\beta^\dagger |n \rangle. \qquad
\end{eqnarray}
In the above expressions, we have taken advantage of the reality of the matrix elements between Fock states, 
$\langle n| f_\beta |m \rangle = \langle m| f_\beta^\dagger |n \rangle$, in order to guarantee
the correct transformation properties of the $R$-operators under the gauge group transformations 
discussed in \sec{sec:gauge}.

At the saddle-point level, on the other hand, when the boson fields are treated as probability
amplitudes and the constraints are satisfied only on average, the
expression of $\underline{d}_\alpha^\dagger$ must be modified\cite{Kotliar_Ruckenstein:86}, in
order to recover the correct normalization of transition
amplitudes in the non-interacting limit. For this purpose, it is easier to define the physical
electron operator in the orbital basis $\{\lambda\}$ in which the
quasiparticle and quasihole density matrices are diagonal,
\begin{eqnarray}
 \hat{\Delta}^{(p)}_{\alpha\beta} & \equiv &
  \sum_{Anm} \phi_{An}^* \phi_{Am}
   \langle m| f_\alpha^\dagger f_{\beta} |n \rangle \nn\\
  & = & \sum_\lambda U_{\alpha\lambda} \xi_\lambda U^\dagger_{\lambda\beta},\\
 \hat{\Delta}^{(h)}_{\alpha\beta} & \equiv &
  \sum_{Anm} \phi_{An}^* \phi_{Am}
   \langle m| f_{\beta} f_\alpha^\dagger |n \rangle \nn\\
  & = & \sum_\lambda U_{\alpha\lambda} (1-\xi_\lambda) U^\dagger_{\lambda\beta},
\end{eqnarray}
where $\xi_\lambda=\sum_{An} |\Omega_{An}|^2 n_\lambda$ is the
probability to find the system in a state such that $n_\lambda=1$,
i.e., with a quasi-particle in the orbital $\lambda$ (note that $\sum_{An}
|\Omega_{An}|^2=1$). In these expressions, the quasiparticle
operators referred to the new orbitals are related to the old ones
by the unitary transformation
\begin{equation}
 f_\alpha^\dagger = \sum_\lambda
 U_{\alpha\lambda} \psi_\lambda^\dagger,
\end{equation}
while the boson fields transform with the corresponding rotation
of the Fock states (summation over repeated indices is implied)
\begin{equation}
 |n\rangle_f = \mathcal{U}(U)_{nm}|m\rangle_\psi,\qquad
 \phi_{An} = \mathcal{U}(U)_{nm} \Omega_{Am}.
\end{equation}
In such basis, the transition amplitude between states with
$n_\lambda=0$ in the initial [final] configuration, and $n_\lambda=1$ in
the final [initial] one, must be normalized by the factor
$1/\sqrt{\xi_\lambda(1-\xi_\lambda)}$, yielding the following expression for
the physical electron operator:
\begin{eqnarray}
 \underline{d}_\alpha^\dagger &=&
 \sum_{AB,\,nm,\,\lambda}
 \frac{\langle A| d_\alpha^\dagger |B\rangle\, \Omega_{An}^* \Omega_{Bm} }%
 {\sqrt{\xi_\lambda[\Omega](1-\xi_\lambda[\Omega])}} \; \times \nn \\
 & & \times \; \left[\langle n| \psi_\lambda^\dagger |m \rangle \psi_\lambda^\dagger +
 \langle n| \psi_\lambda |m \rangle \psi_\lambda \right].
\end{eqnarray}
Rotating back to the original basis, we finally get, for the saddle-point expressions of the $R$-matrices,
\begin{equation}
\begin{array}{rcl}
 R^{(p)}_{\alpha\beta}[\phi]^* &=& \ds \sum_{AB,\,nm,\,\gamma}
 \langle A| d_\alpha^\dagger |B\rangle\, \phi_{An}^* \phi_{Bm} 
 \langle m| f_\gamma |n \rangle M_{\gamma\beta}, \qquad \\
 R^{(h)}_{\alpha\beta}[\phi] &=& \ds \sum_{AB,\,nm,\,\gamma}
 \langle A| d_\alpha^\dagger |B\rangle\, \phi_{An}^* \phi_{Bm} 
 \langle m| f_\gamma^\dagger |n \rangle M_{\beta\gamma}, \qquad
\end{array}
\end{equation}
where
\begin{equation}
 M_{\gamma\beta}
 =\left[\frac{1}{2}\left(\hat{\Delta}^{(p)}\hat{\Delta}^{(h)} +
 \hat{\Delta}^{(h)}\hat{\Delta}^{(p)}\right)\right]^{-\frac{1}{2}}_{\gamma\beta}
\end{equation}
is the particle-hole symmetrized version of the normalization
factor, expressed in the original basis.

\subsection{Functional integral representation}
\label{sec:gauge}

The partition function of a generic multi-orbital Hamiltonian \eq{eq:generic_Ham} can be 
formally written, in terms of auxiliary fields, as
\begin{equation}
\Z = \int\! \D[f,f^\dagger] \, \D[\{\phi\},\{\A \}] \, e^{ -\int_0^\beta\! d\tau \mathcal{L}(\tau) }, \label{eq:partition_funct}
\end{equation}
where we have introduced, along with slave bosons and auxiliary fermions,
a set of Lagrange multiplier fields $\{\A_i(\tau)\}$ that allow to enforce, 
at each lattice site $i$ and imaginary-time value $\tau$, the constraints \eqs{eq:const1}{eq:const3}.
The Lagrangian functional entering the above expression reads
\begin{eqnarray} 
\mathcal{L}(\tau) &=& \sum_i \left( \phi_{An,i}^\dagger \detau \phi_{An,i}
 + f_{\alpha,i}^\dagger \detau f_{\alpha,i} \right. \nn \\
 & & \left. +\, \underline{H}^{\ph{\dagger}}_\mathrm{const}[i] + \underline{H}_\mathrm{loc}[i]  \right) + \underline{H}_\mathrm{kin}
\label{eq:Lagrangian1}
\end{eqnarray}
(except for lattice sites, summation over repeated indices is always implied throughout this section),
where $\underline{H}_\mathrm{loc}$ and $\underline{H}_\mathrm{kin}$ are, respectively, 
the representatives of the local and kinetic part of the Hamiltonian \eq{eq:generic_Ham}
in the enlarged Hilbert space, while $\underline{H}_\mathrm{const}$ contains the constraint interactions between 
auxiliary fields and Lagrange multipliers. 

In order to derive the expressions of the Hamiltonian terms in \eq{eq:Lagrangian1}, 
and thereby identify the underlying symmetry group of the Lagrangian, it is convenient to collect all 
the local fermionic degrees of freedom (either physical or auxiliary) into a
$2M$-component Nambu-Gor'kov spinor:
$$ 
\Xi_{i} \equiv
\left(\begin{array}{c} \{d_{\alpha,i}\} \\ \{d^\dagger_{\alpha,i}\} \end{array}\right), \qquad
\Psi_{i} \equiv
\left(\begin{array}{c} \{f_{\alpha,i}\} \\ \{f^\dagger_{\alpha,i}\} \end{array}\right). 
$$   
In such formalism, the representation of physical electrons in terms of bosons and quasiparticles \eq{eq:R-matrices} is 
simply written as $\underline{\Xi}_i = \RR_i \Psi_i$, where the local $2M\times 2M$ matrix operator
\begin{equation}
 \RR_i \equiv \RR[\phi_i] = \left(%
\begin{array}{cc}
  \RRp[\phi_i] & \RRh[\phi_i]^* \\
  \RRh[\phi_i] & \RRp[\phi_i]^* \\
\end{array}%
\right) \label{eq:RR-matrix}
\end{equation}
is defined in terms of the boson fields $\phi_{An,i}$ associated to the corresponding site $i$
(note that, to lighten the notation, site indices were omitted in previous sections).
The representation of the kinetic Hamiltonian is then readily obtained as
\begin{eqnarray}
 \underline{H}_{\rm kin} &=&  \sum_{ij} 
 t^{\alpha\beta}_{ij} \underline{d}_{\alpha,i}^\dagger \underline{d}_{\beta,j} \nn \\
 &=& \frac{1}{2} \sum_{ij} \underline{\Xi}^\dagger_i \, \tilde{\bf t}_{ij} \, \underline{\Xi}_j \nn \\
 &=& \frac{1}{2} \sum_{ij} \Psi^\dagger_i \, \RR^\dagger_i \tilde{\bf t}_{ij} \RR_j \, \Psi_j , \label{eq:H_kin_bosons} \\
 \tilde{\bf t}_{ij} &=& \left( 
 \begin{array}{cc}
  {\bf t}_{ij} & 0 \\
  0 & -{\bf t}_{ij}^* \\
 \end{array} \right), 
\end{eqnarray}
${\bf t}_{ij}$ being the real-space hopping matrix, whose Fourier transform gives
the band dispersion matrix $\ee(\kk)$ defined in \ceq{eq:H_kin_gen}.

On the other hand, the local terms of the model Hamiltonian, which may include any kind of on-site interaction 
between (physical) electrons, are represented, within the enlarged Hilbert space, by a purely quadratic boson Hamiltonian:
\begin{equation}
 \underline{H}_\mathrm{loc}[i] = \langle A| H_\mathrm{loc} |B\rangle \phi_{An,i}^\dagger \phi_{Bn,i} 
 = E_\Gamma \phi_{\Gamma n,i}^\dagger \phi_{\Gamma n,i}, \label{eq:H_loc_bosons}
\end{equation}
where the physical states $\{|\Gamma \rangle\}$ denote the eigenstates of $H_{\rm loc}$.

\subsubsection*{Gauge invariance}

In order to discuss the symmetry structure of the Lagrangian \eq{eq:Lagrangian1}, 
we begin to notice that the auxiliary-fields Hamiltonian $\underline{H} = \underline{H}_{\rm kin}
+ \sum_i \underline{H}_{\rm loc}[i]$ is invariant under the following gauge transformations:
\begin{eqnarray}
\Psi_i(\tau) & \ra & \UU_i(\tau) \Psi_i(\tau), \label{eq:gauge_psi}\\
\phi_{An,i}(\tau) & \ra & e^{i\xi^0_i(\tau)} \times \mathcal{U}[\UU]_{nn'} \phi_{An',i}(\tau). \label{eq:gauge_bosons} 
\end{eqnarray}
The unitary matrix $\UU_i(\tau)$ in \eq{eq:gauge_psi} represents an arbitrary $SO(2M)$ rotation of
quasiparticle operators acting independently on each site, and it is conveniently parameterized as
\begin{equation}
\UU_i(\tau) = e^{i \xi_i^a(\tau) \TT^a}, \label{eq:U_nambu_exp}
\end{equation}
the $\TT^a$ matrices being a $2M$-dimensional representation of the $M(2M-1)$ group generators.
We note, however, that such matrices are not expressed in the usual form of an
orthogonal group generator (namely, as purely imaginary antisymmetric matrices), but are instead of the form
\begin{equation}
\TT^a = \left( \begin{array}{cc}
  \TT^a_{\rm H} & \TT^a_{\rm A} \\
  -{(\TT^a_{\rm A})}^* & -{(\TT^a_{\rm H})}^* 
  \end{array} \right),
\end{equation}
with $\TT^a_{\rm H}$ and $\TT^a_{\rm A}$ denoting, respectively, $M\times M$ Hermitian and antisymmetric matrices.
In other words, the $2M$-dimensional Nambu spinors do not transform with real orthogonal matrices under $SO(2M)$ rotations,
even though they clearly form a real representation of the gauge group: 
\begin{eqnarray}
\Psi_i^\dagger &=& \Psi_i^T {\cal E}, \label{eq:real_nambu}\\
\UU_i^* &=& {\cal E} \UU_i {\cal E}, \\
{\cal E} &=& \left(\begin{array}{cc}
  0 & \ident \\
  \ident &  0 \\ 
  \end{array}\right).
\end{eqnarray}
The transformation law of the boson fields, on the other hand,
is characterized by the unitary transformation of Fock states
$
|n\rangle \ra  \mathcal{U}[\UU]_{nn'} |n'\rangle
$
that is associated to the $SO(2M)$ rotation of quasiparticle operators,
plus an additional $U(1)$ factor $e^{i\xi^0_i(\tau)}$ under which 
the Nambu spinors are neutral. Following the exponential parameterization 
of $\UU_i(\tau)$, we can then similarly write
\begin{equation}
\mathcal{U}[\UU]_{nn'} = \left[e^{i \xi_i^a(\tau) \J^a}\right]_{nn'},  \label{eq:U_bosons_exp}
\end{equation}
where $\J^a_{nn'}$ are the $SO(2M)$ group generators expressed in the Fock space representation.

While the gauge invariance of $\underline{H}_{\rm loc}[i]$  follows 
immediately from its definition \eq{eq:H_loc_bosons}, in the case of $\underline{H}_{\rm kin}$ we still need to
establish the transformation properties of the $\RR_i$ operators, defined 
in \ceq{eq:RR-matrix}. These, however, can be readily obtained by writing such operators 
directly in terms of the physical and auxiliary Nambu spinors, in matrix notation: 
\begin{eqnarray}
 \RR_i & = & \frac{1}{M} 
 \langle A| \Xi |B\rangle\, \phi_{An,i}^\dagger \phi_{Bm,i} \langle m| \Psi^\dagger |n \rangle, \\
 \RR_i(\tau) & \ra & \RR_i(\tau) \UU_i^\dagger(\tau).
\end{eqnarray}
The above transformation law ensures the gauge invariance of the physical electron operator
$\underline{\Xi}_i = \RR_i \Psi_i$, and hence of the whole kinetic Hamiltonian \eq{eq:H_kin_bosons}.  

In the discussion of the gauge invariance of the Lagrangian \eq{eq:Lagrangian1}, we are thus left with the 
time-derivatives and constraints terms, whose transformation properties are closely related   
to each other. The time-derivative terms, in fact, are clearly not invariant under the transformations
\eq{eq:gauge_psi} and \eq{eq:gauge_bosons}, which generate inhomogeneous terms proportional to the
time derivatives of the rotation parameters (e.g., $\detau \xi_i^a$ and $\detau \xi_i^0$). 
Such terms, however, can be reabsorbed by a corresponding inhomogeneous transformation of the Lagrange multiplier 
fields\cite{Fresard-Wolfle:92,Fresard-Kopp:01-07} 
(which may be regarded as ``gauge bosons''), making the whole Lagrangian gauge invariant. 

To show how this mechanism works, we rewrite the $M^2 + M(M-1)$ local constraints \eq{eq:const2} and \eq{eq:const3}
in the following way:
\begin{equation}
 \phi_{An,i}^\dagger \phi_{An',i}
 \langle n'| \Psi^\dagger \TT^a \Psi |n \rangle = \Psi_i^\dagger \TT^a \Psi_i,
\end{equation}
where we have made use of the ``orthogonality'' between the $SO(2M)$ $\TT^a$ matrices. 
It is then straightforward (though somewhat lengthy) to verify that the Fock-space generators $\J^a_{nn'}$, introduced
in \ceq{eq:U_bosons_exp}, are represented by 
\begin{equation}
\J^a_{nn'} = -\frac{1}{2} \langle n'| \Psi^\dagger \TT^a \Psi |n \rangle
\end{equation}
(in other words, the above matrices provide a faithful representation of the $SO(2M)$ Lie algebra),
so that we can finally write:
\begin{eqnarray}
 \underline{H}_\mathrm{const}[i] 
 &=& \A_i^a \left( \frac{1}{2} \Psi_i^\dagger \TT^a \Psi_i + \phi_{An,i}^\dagger \J^a_{nn'} \phi_{An',i} \right) \nn \\
 & & +\, \A_i^0 \left( \phi_{An,i}^\dagger \phi_{An,i} -1 \right).
\end{eqnarray}
Together with the time-derivative terms, the above interactions may be arranged in 
``covariant derivatives'' acting on the auxiliary fields (fermions and bosons),
\begin{eqnarray}
 \frac{1}{2} \Psi_i^\dagger D_\tau \Psi_i & = &  \frac{1}{2} \Psi_i^\dagger \left[ \detau + \A_i^a \TT^a \right] \Psi_i ,\\
 \phi_{An,i}^\dagger D_\tau \phi_{An,i} & = & \phi_{An,i}^\dagger 
 \! \left[ \left(\detau + \A_i^0 \right)\! \delta_{nn'} + \A_i^a \J^a_{nn'} \right]\! \phi_{An',i}, \nn \\
\end{eqnarray} 
where the role of gauge fields is played by the Lagrange multipliers $\A_i^a$ and $\A_i^0$.
It is then easily checked that the Lagrangian 
\begin{equation} 
\mathcal{L} = \! \sum_i \! \left[ \phi_{An,i}^\dagger D_\tau \phi_{An,i}
 + \frac{1}{2} \Psi_i^\dagger D_\tau \Psi_i 
 + \underline{H}_\mathrm{loc}[i] - \A_i^0 \right] + \underline{H}_\mathrm{kin}
\end{equation}
is indeed invariant under the $SO(2M)\otimes U(1)$ gauge group, provided the Lagrange multipliers transform
as gauge fields in the adjoint representation\cite{noteA0}:
\begin{eqnarray}
 \A_i^0 & \ra & \A_i^0 - i \detau \xi_i^0(\tau) , \\
 \A_i^a \TT^a & \ra & \UU_i(\tau) \left[ \A_i^a \TT^a + \detau \right] \UU_i^\dagger(\tau). \label{eq:trans_Lag}
\end{eqnarray}
Note that the transformation law of $\A_i^a \TT^a$ induces a corresponding transformation
of $\A_i^a \J^a_{nn'}$ that has exactly the same structure of \eq{eq:trans_Lag}, with
the $\UU_i$ matrix replaced by $\U[\UU]_{nn'}$. More precisely, both transformation laws
descend from that of the Lagrange multiplier field $\A_i^a(\tau)$, 
which for infinitesimal rotations transforms as
\begin{equation}
 \A_i^a  \ra  \A_i^a + f^{abc} \A_i^b \xi_i^c  - i \detau \xi_i^a + \ord(\xi^2),
\end{equation}
$f^{abc}$ being the structure constants of the group.

Finally, it is worth mentioning that for $M=2$ (single-band models) the orthogonal group
$SO(4)$ is locally isomorphic to $SU(2)\otimes SU(2)$,
so that the gauge group structure of the present formalism reduces to
that of the spin-charge invariant formalism of Ref.~\onlinecite{Fresard-Wolfle:92}.

\subsubsection*{Gauge fixing}

As discussed in Ref.~\onlinecite{Fresard-Wolfle:92}, 
the gauge invariance of the functional integral representation causes 
\ceq{eq:partition_funct} to contain integration over spurious degrees of freedom,
namely the physically equivalent field configurations connected to each other by gauge group trajectories.
It is thus necessary to impose a ``gauge fixing'' condition that removes the integration over the unphysical 
degrees of freedom, as it is usually done in gauge field theories\cite{Faddeev-Popov}.

We choose to work in the so-called ``radial gauge''\cite{radial-gauge}, in which the complex boson fields
are represented through real amplitudes and complex phase fields. In this representation, the $SO(2M)\otimes U(1)$ gauge transformations
allow to remove $M(2M-1) +1$ phase variables, so that a corresponding number of boson fields can be reduced to purely real amplitudes, 
with no phase fluctuations. The boson fields that remain fully complex, on the other hand,
continue to display some phase dynamics, which is responsible for the incoherent features of the spectrum\cite{Jolicoeur-Le_Guillou:91} 
(e.g., lower and upper Hubbard bands).  

It is beyond our purpose to enter into the formal details of the radial gauge representation, thoroughly derived in
Ref.~\onlinecite{Fresard-Kopp:01-07}. Nevertheless, it is worthwhile to observe, here, that such gauge fixing procedure allows to avoid 
Elitzur's theorem\cite{Elitzur}, which would prevent the slave bosons to acquire a non-zero expectation value,
making therefore legitimate the use of the saddle-point approximation.

\subsection{Saddle-point approximation}
\label{sec:saddle-point} 

The saddle-point approximation of the functional integral \eq{eq:partition_funct} is 
obtained by considering the slave bosons and Lagrange multipliers as static variables, 
corresponding to the time-averages of these fields. Moreover, we will assume a homogeneous
spatial structure of the saddle-point solution, so that we can finally set: $\phi_{An,i}(\tau) \mapsto \varphi_{An}$, 
$\A_i^a(\tau) \mapsto \A^a$ and $\A_i^0(\tau) \mapsto \A^0$.

In such approximation, all the bosonic amplitudes $\varphi_{An}$ are assumed to have a constant phase, 
in contrast to the radial gauge representation, where we are allowed to remove (fix) 
a limited number of complex phases (namely $M(2M-1) +1$). The phase fluctuations of those fields that 
remain intrinsically complex are thus ruled out, precluding the possibility of describing the high-energy physics of a given model. 
The low-energy features, on the other hand, will be suitably described in terms of coherent Landau-Bogoliubov quasiparticles,
providing a Fermi-liquid description of metallic and superconducting states.

At the saddle point, the free energy per site $\Omega=-(1/\beta N) \ln \Z$ is obtained as
the (minimum) stationary value of the following free-energy functional:
\begin{eqnarray}
\Omega[\{\varphi\},\{ \A\}] &=&  - \frac{1}{\beta N} \ln \Z_f -\A^0  \label{eq:free_en}  \\
 & & + \sum_{AB\,nm} \varphi_{An}^* \left[ \langle A| H_\mathrm{loc} |B\rangle \delta_{nm}^{\ph{0}} \right. \nn \\
 & & + \left. \left( \A^0 \delta_{nm} + \A^a \J^a_{nm} \right) \delta_{AB} \right] \varphi_{Bm}, \nn
\end{eqnarray}
where $N$ is the total number of sites and $\Z_f$ represents
the Gaussian integral over the auxiliary fermions,
\begin{eqnarray}
\Z_f & = & 
  \int\! \D \Psi \exp\! \left[ -\frac{1}{2} \!  \int_0^\beta \!\!\! d\tau \sum_{ij}
 \Psi_i^\dagger \! \left[ \delta_{ij} D_\tau  + \RR^\dagger \tilde{\bf t}_{ij} \RR \right] \! \Psi_j \right] \!\! \nn \\
 & = & \int\! \D \Psi \exp\! \left[ -\frac{1}{2} \!  \int_0^\beta \!\!\! d\tau \sum_{\kk }
 \Psi_{\kk}^\dagger \left[ \detau + \mathbf{h}(\kk) \right] \Psi_{\kk} \right]\! . \label{eq:partition_fermions}
\end{eqnarray}
In the last expression, $\mathbf{h}(\kk)$ denotes the momentum-space quasiparticle energy matrix: 
\begin{equation}
\mathbf{h}(\kk) = \RR^\dagger[\varphi] \left( \begin{array}{cc}
  \ee(\kk) & 0 \\
  0 & -\ee(-\kk)^* \\
\end{array} \right) \RR[\varphi] + \A^a \TT^a. \label{eq:QP_h_matrix}
\end{equation}
To evaluate the functional integral in \ceq{eq:partition_fermions}, we note that, up to irrelevant constants,
$\Z_f^2=\det\left[\detau + \mathbf{h}(\kk)\right]$, so that we can write
\begin{eqnarray}
 \ln \Z_f &=& \frac{1}{2} \ln \det \left[\detau + \mathbf{h}(\kk)\right] \nn\\
 &=& \frac{1}{2} \sum_\kk \mathrm{tr} \ln \left[1 + e^{-\beta\mathbf{h}(\kk)}\right].
\end{eqnarray}

The saddle-point equations are then obtained by setting to zero all the
partial derivatives of the free-energy functional \eq{eq:free_en} with respect to the
slave boson amplitudes and Lagrange multipliers. 
For practical calculations, however, it is useful to consider a different basis set 
for the $M(2M-1)$ Lagrange multipliers $\A^a$, belonging the adjoint representation of $SO(2M)$. 
In place of them, in fact, we can equivalently use 
the set of independent matrix elements of $\A^a \TT^a$, parameterized as follows:
\begin{eqnarray}
 \A^a \TT^a &=& \left( \begin{array}{cc}
   {\bf \Lambda} & {\bf \Pi} \\
   -{\bf \Pi}^* & -{\bf \Lambda}^* 
   \end{array} \right), \\
 \Lambda_{\alpha \beta} &=& \Lambda_{\beta \alpha }^*, \nn \\ 
 \Pi_{\alpha \beta} &=& -\Pi_{\beta \alpha }. \nn
\end{eqnarray}
With this choice, the saddle-point equations can be derived without 
knowing the explicit expressions of the $\TT^a$ matrices, differentiating the free-energy
functional directly with respect to $\Lambda_{\alpha \beta}$ and $\Pi_{\alpha \beta}$.
To this end, we note that $\Omega[\{\varphi\},\{ \A\}]$ can be easily expressed in terms of the new set of Lagrange multipliers 
by means of the following relation:
\begin{eqnarray}
\A^a \J^a_{nn'} & = & -\frac{1}{2} \langle n'| \Psi^\dagger (\A^a \TT^a) \Psi |n \rangle \nn \\
 & = & -\frac{1}{2} \left[ \Lambda_{\alpha \beta}  
        \langle n'| \left(f_\alpha^\dagger f_{\beta} - f_{\beta}f_\alpha^\dagger \right)|n \rangle 
        \ph{f_{\beta}^\dagger}  \right.  \\
 & &  + \left. \Pi_{\alpha\beta}  \langle n'| f_\alpha^\dagger f_{\beta}^\dagger |n \rangle
      + \Pi_{\alpha\beta}^*  \langle n'| f_{\beta} f_\alpha |n \rangle \right]. \nn
\end{eqnarray}

\subsection{Green's functions and observables}
\label{sec:Green_fun}

After solving the saddle-point equations, we can finally obtain
the expressions for the quasiparticle and physical electron
propagators, conveniently written here in Nambu notation. 
For quasiparticles, the propagator is defined as
\begin{eqnarray}
 \mathbf{D}_f(\kk,\tau) &=& -\langle T \Psi_\kk(\tau) \Psi_\kk^\dagger(0)\rangle  \label{eq:Df_propag} \\[0.2cm]
 &=& \left(%
\begin{array}{cc}
  \mathbf{G}_f(\kk,\tau) & {\mathbf{F}_f^\dagger(\kk,-\tau)}_{\phantom{\int}} \\
  \mathbf{F}_f(\kk,\tau) & -\mathbf{G}_f^T(-\kk,-\tau)\\
\end{array}%
\right),  \nn
\end{eqnarray}
where $G_f(\kk,\tau)_{\alpha\beta}=-\langle T f_{\kk\alpha}(\tau)
f_{\kk\beta}^\dagger(0)\rangle$ and
$F_f(\kk,\tau)_{\alpha\beta}=-\langle T
f_{-\kk\alpha}^\dagger(\tau) f_{\kk\beta}^\dagger(0)\rangle$ are the normal and anomalous quasiparticle Green's functions. 
Following \ceq{eq:partition_fermions}, we can then readily write the quasiparticle inverse propagator as 
\begin{equation}
 \mathbf{D}_f^{-1}(\kk,\omega) = \omega - \mathbf{h}(\kk). \label{eq:Df_h}
\end{equation}
The physical electron propagator, on the other hand, is 
defined by
\begin{equation}
 \mathbf{D}_d(\kk,\tau) = -\langle T \Xi_\kk(\tau) \Xi_\kk^\dagger(0)\rangle,
\end{equation}
where $\Xi_\kk^\dagger \equiv
\left(\{d_{\kk\alpha}^\dagger\},\{d_{-\kk\alpha}\}\right)$ is the Nambu spinor
containing the physical degrees of freedom, represented in terms of slave boson amplitudes and quasiparticles
by $ \underline{\Xi}_\kk = \RR[\varphi] \Psi_\kk$. 
The expression for the inverse physical propagator is thus written as
\begin{equation}
 \mathbf{D}_d^{-1}(\kk,\omega) = [\RR^\dagger]^{-1}\left[\omega - \mathbf{h}(\kk)\right] \RR^{-1}.
\end{equation}
Using the corresponding expression for the ``bare''
physical propagator, 
\begin{eqnarray}
 \mathbf{D}_{d0}^{-1}(\kk,\omega) &=& \omega - \left[\tilde{\ee}_0 + \tilde{\ee}(\kk)\right],\\
 \tilde{\ee}_0 + \tilde{\ee}(\kk) &=& \left(%
\begin{array}{cc}
  \ee_0 + \ee(\kk) & 0 \\
  0 & -\left[\ee_0 + \ee(-\kk)\right]^* \\
\end{array}%
\right)
\end{eqnarray}
($\ee_0$ represents the one-body part of $H_\mathrm{loc}$),
we can then find the saddle-point approximation for the self-energy:
\begin{eqnarray}
 \mathbf{\Sigma}_d(\omega) &=& \mathbf{D}_{d0}^{-1} - \mathbf{D}_d^{-1} \nn \\
 &=& \omega\left(1-[\RR\RR^\dagger]^{-1}\right) - \tilde{\ee}_0 \nn  \\
 & & +\, [\RR^\dagger]^{-1} \left( \begin{array}{cc}
       {\bf \Lambda} & {\bf \Pi} \\
       -{\bf \Pi}^* & -{\bf \Lambda}^* 
       \end{array} \right) \RR^{-1}.  
\end{eqnarray}
The cancellation of the $k$-dependence, in the above equation, follows from the definition 
of the QP energy matrix, \ceq{eq:QP_h_matrix}. Indeed, this form of the self-energy
is just the one we would expect from the saddle-point (mean-field)
approximation, which freezes spatial and dynamical fluctuations.
From the linear term in $\omega$, one readily obtains the matrix of
quasiparticle spectral weights: 
\begin{eqnarray}
\mathbf{Z} & = & \left[ 1 - \frac{\de {\bf \Sigma}_d}{\de \omega} \right]^{-1}_{\omega=0}\nn \\
 & = & \RR\RR^\dagger.
\end{eqnarray}

We conclude this formal section by writing the representation, 
in the enlarged Hilbert space, of the
local physical density operator, whose 
expectation value defines the average number of electrons per site:
\begin{equation}
 \underline{\hat{n}}^{(d)} = \sum_{\alpha} \underline{d}_{\alpha}^\dagger \underline{d}_{\alpha}
 = \sum_A N_A \sum_n \phi_{An}^\dagger \phi_{An}.
\end{equation}
As mentioned previously, we remark that this expression differs substantially 
from that for the local quasiparticle density,
\begin{equation}
 \sum_{\alpha} f_{\alpha}^\dagger f_{\alpha}
 = \sum_{An} \sum_\alpha  n_\alpha\, \phi_{An}^\dagger \phi_{An} ,
\end{equation}
which is not a physical quantity and depends from the choice of
the QP basis set. More generally, the representation of any (local) two-particle physical operator  
may be easily obtained in terms of boson 
operators only: for particle-hole operators we find
\begin{equation}
 \underline{d}_\alpha^\dagger \underline{d}_\beta \,=\, \sum_{AB}
   \langle A| d_\alpha^\dagger d_\beta |B\rangle \sum_n \phi_{An}^\dagger \phi_{Bn},
\end{equation}
while for particle-particle operators we have 
\begin{equation}
 \underline{d}_\alpha^\dagger \underline{d}_\beta^\dagger \,=\, \sum_{AB}
   \langle A| d_\alpha^\dagger d_\beta^\dagger |B\rangle \sum_n \phi_{An}^\dagger \phi_{Bn}. \label{eq:part-part_bos}
\end{equation}

\section{Application to the three-orbital model for fullerides}\label{sec:fullerides}

\subsection{The model}\label{sec:model}

In this section we present an explicit application of the rotationally invariant slave boson approach. To this aim we have chosen a three-orbital Hubbard-like model that has been used to understand the role of strong correlations in alkali-doped fullerides. 
The physics of alkali-doped fullerene systems $\mathrm{A_n C_{60}}$ represents 
an optimal playground in understanding the key role of strong correlations on high-temperature superconductors: 
indeed, these systems display a relatively high \tc compared to ordinary Bardeen-Cooper-Schrieffer (BCS) superconductors, and, similarly to
what is found in cuprate systems, the enhancement of \tc seems to be closely related to the proximity of a Mott insulating 
phase\cite{Capone-Science,CaponeRMP,NatureMaterialsvari}. 
Although the nature of the pairing mechanism is most likely
different from that characterizing cuprate superconductors, in fullerene systems being due to ordinary electron-phonon (vibron) interaction\cite{Gunnarsson:RMP}, and in spite of the different symmetry of the order parameter (\sw in fullerides), the phase 
diagram as a function of the inter-molecule separation in these systems presents strong similarities to that of cuprates as a function of doping\cite{NatureMaterialsvari}, providing an independent example of the key role of electronic correlations in enhancing superconductivity.  

The local Hamiltonian describing the $\mathrm{{C_{60}}^{n-}}$ molecular ion, assuming
rotational invariance within the threefold degenerate level $t_{1u}$ hosting the valence electrons provided by the alkali metals, can be written, for a generic site $i$, as
\begin{eqnarray}
H_\mathrm{loc}[i] &=& 
 \frac{U}{2} \hat{n}_i^2 
 - J_\mathrm{H} \left[ 2\, {\mathbf{S}}_i \cdot {\mathbf{S}}_i + \frac{1}{2} {\mathbf{L}}_i \cdot {\mathbf{L}}_i 
 + \frac{5}{6} (\hat{n}_i -3)^2 \right] \nn \\[0.2cm]
 & & + H_\mathrm{JT}, \label{eq:H_loc_JT}
\end{eqnarray} 
where the three terms represent, respectively, the global on-site Coulomb repulsion, Hund's rules splitting ($J_\mathrm{H} > 0$)
and the Jahn-Teller coupling between electrons and the vibrational modes (vibrons) of $\mathrm{C_{60}}$.  
In the above expression, $\hat{n}_i = \sum_{a\sigma}d^\dagger_{i,a\sigma} d_{i,a\sigma}$ 
is the local electron number operator ($a=1,2,3$ labels the $t_{1u}$ orbitals and $\sigma = \uparrow,\downarrow$ the spin components), 
while ${\mathbf{S}}_i$ and ${\mathbf{L}}_i$ are, respectively, the local spin and orbital angular-momentum operators,
\begin{eqnarray}
{\mathbf{S}}_i &=& \frac{1}{2} \sum_{a,\,\sigma\sigma'} d^\dagger_{i,a\sigma} \hat{\bm{\sigma}}_{\sigma\sigma'} d_{i,a\sigma'}, \\
{\mathbf{L}}_i &=& \sum_{ab,\,\sigma} d^\dagger_{i,a\sigma} \hat{\bm{\ell}}_{ab} d_{i,b\sigma},
\end{eqnarray}
where $\hat{\bm{\sigma}}_{\sigma\sigma'}$ are the Pauli matrices and $\hat{\bm{\ell}}^{(a)}_{bc} = i \eps_{bac}$ 
the $O(3)$ group generators characterizing orbital rotations. 

The Jahn-Teller Hamiltonian involves both electron and vibron field
operators, but if we are interested only in the electron dynamics we can formally integrate out the vibronic degrees of freedom, 
obtaining an effective action for the electron operators only. If performed exactly, this procedure would clearly 
generate non-instantaneous (i.e., time-dependent) interaction terms, preventing a purely electronic Hamiltonian formulation of the effective action; 
however, if we assume the vibronic frequencies to be much higher than the relevant electronic scales, we can take the 
anti-adiabatic limit of the electron effective action, neglecting retardation effects and considering an instantaneous interaction term
which preserves the symmetries of the original local Hamiltonian. While this approximation may be questionable for 
fullerides, it will not affect our claims since the neglect of retardation can only disfavor superconductivity, and analogous results have been obtained in a similar model that takes into account the phonon dynamics\cite{Han:03}.
The Jahn-Teller interaction can then be reabsorbed in Hund's term, 
and the resulting Hamiltonian is simply given by the first two terms of \ceq{eq:H_loc_JT} with a renormalized
Hund's coupling $J_\mathrm{H} \mapsto -J = J_\mathrm{H} - \frac{3}{4} E_\mathrm{JT}  <0$, $E_\mathrm{JT}$
being the characteristic Jahn-Teller energy gain\cite{note4}.
The net effect of the electron-vibron coupling is therefore that of reversing Hund's rules, favoring atomic configurations with low 
spin and orbital angular momentum. The inversion of the Hund's rule is experimentally confirmed by the low-spin state of both tetravalent\cite{A4C60} and trivalent\cite{spinhalf} fullerides and by the presence of a spin gap\cite{spingap}.

Given the local Hamiltonian for the $\mathrm{{C_{60}}^{n-}}$ molecular ion, the expression for a tight-binding
electronic Hamiltonian describing $\mathrm{A_n C_{60}}$ solids will then read
\begin{equation}
\label{finalham}
 H_\mathrm{latt} = \sum_{ij,\,ab,\,\sigma} t^{ab}_{ij}\,
 d^\dagger_{i,a\sigma} d_{j,b\sigma} + \sum_i H_\mathrm{loc}[i] - \mu \sum_i \hat{n}_i,
\end{equation}
where $t^{ab}_{ij}$ are the inter-site hopping amplitudes (including possible inter-band hybridization terms) 
and $\mu$ is the chemical potential controlling the average electron density. 
We should note, however, that inter-band hybridization can actually be avoided 
whenever the hopping terms are restricted to nearest-neighbors only, $t^{ab}_{ij} = t^{ab} \delta_{\langle ij \rangle}$; 
in such case, indeed, we can exploit the $O(3)$ orbital symmetry of $H_\mathrm{loc}$ in order to
diagonalize $t^{ab}$, so that we can set, without loss of generality, $t^{ab}_{ij} = t^{a} \delta_{ab} \delta_{\langle ij \rangle}$.
Throughout our analysis we will use the latter expression for the hopping matrix elements, focusing in particular
on the orbitally degenerate case $t^{a} = -t$ (we will consider the possibility of a crystal-field splitting of the three bands
and of different bandwidths in a future study).

\subsection{Slave-boson representation of the model}

As pointed out in previous studies\cite{A4C60,CaponePRL,Capone-Science,CaponeRMP}, 
the zero-temperature phase diagram of the model (\ref{finalham}) as a function 
of the ratios $J/W$ and $U/W$ ($W$ being the non-interacting bandwidth) displays
several interesting features, the most striking one being undoubtedly the presence of a strongly enhanced
superconducting phase in the proximity of the metal-to-insulator Mott transition.
The model represents therefore a valuable test for the rotationally invariant slave boson method, at the same time
providing an analytical tool to treat strongly correlated superconductors.

\subsubsection*{Slave-boson amplitudes}

From the rotationally-invariant slave-boson representation of physical states defined by 
\ceq{eq:phys_ket}, it should be clear that there are, in principle, $2^{2M-1}$ (with $M = 6$ in our model) slave-boson 
fields $\phi_{An}$ describing the system. However, we must note that when the partition function is 
approximated at saddle-point level, i.e., when the slave-boson fields are replaced by 
their mean-field expectation values ($\phi_{An} \ra \langle \phi_{An} \rangle \equiv \varphi_{An}$),
the number of independent slave-boson amplitudes entering the saddle-point equations becomes much smaller, as
we will show in the following, its specific
value depending on which symmetries of the model Hamiltonian remain unbroken.

As discussed at end of \sec{sec:model}, we choose the hopping matrix to be diagonal 
and degenerate in both spin and orbital indices, 
so that the atomic $SU(2)\otimes O(3)$ symmetry characterizing $H_\mathrm{loc}$ can be promoted to a global symmetry 
for the full lattice Hamiltonian $H_\mathrm{latt}$. If we impose this symmetry to be preserved at saddle-point level,
i.e., we do not allow for any spin and orbital ordering, we must then set to zero all the possible order parameters 
which are not invariant under $SU(2)\otimes O(3)$, and this will strongly limit the possibility of independent slave-boson
amplitudes. Indeed, considering the normal state solution of the saddle-point equations, i.e., do not allowing
for a superconducting order parameter, it is quite straightforward to prove, using Wigner-Eckart's theorem, that the non-zero
slave-boson amplitudes must be of the form
\begin{equation}
\varphi_{\Gamma n} = \langle n | \Gamma \rangle \, \Phi (E_\Gamma), \label{eq:norm_amp}
\end{equation}
where we have taken as the basis set for the local physical states the eigenstates 
$\{|\Gamma\rangle\}$ of $H_\mathrm{loc}$, with eigenvalues $E_\Gamma$. Note that in this 
case the quasi-particle Fock states $|n\rangle$ have exactly the same number of particles of 
the physical state $|\Gamma\rangle$ to which they are linked, assuring the solution 
to represent a normal state; indeed, as long as the latter condition is satisfied,
no superconducting order parameter can be ever developed, as can be easily seen taking the expectation values of \ceq{eq:part-part_bos}, 
\begin{equation}
\langle d_\alpha^\dagger d_\beta^\dagger \rangle \,=\, \sum_{\Gamma \Gamma'}
   \langle \Gamma| d_\alpha^\dagger d_\beta^\dagger |\Gamma' \rangle \sum_n \varphi_{\Gamma n}^* \varphi_{\Gamma' n}.  \label{eq:expect_part-part_bos}
\end{equation}  

On the other hand, if we do allow for 
a superconducting symmetry-breaking, we must add to the normal amplitudes defined in \ceq{eq:norm_amp} 
also those amplitudes connecting physical states to QP states with a different number of particles, 
so that particle number would no longer be conserved; however, if we still want to 
preserve the $SU(2)\otimes O(3)$ symmetry as in the normal case, we should consider only those amplitudes which correspond to
an invariant, with respect to spin and orbital rotations, superconducting order parameter. Assuming
pairing to be purely local, corresponding to an $s$--wave order parameter,
the only invariant pairing amplitude is then given by the spin and orbital singlet 
channel
\begin{equation}
\psi^0_{sc} \equiv \left\langle \ts \sum_a  d^\dagger_{i,a\uparrow} d^\dagger_{i,a\downarrow} \right \rangle . \label{eq:sc_order_param}
\end{equation}
At this point, it is worthwhile to observe that \eq{eq:sc_order_param} represents the
most favorable pairing channel even if we do not explicitly impose the $SU(2)\otimes O(3)$ symmetry,
since the local pairing attraction is driven by the reversed Hund's term, which favors the formation of two-particle states locked in the $L=S=0$
spin-orbital configuration. Our assumption of preserving the $SU(2)\otimes O(3)$ symmetry is then fully justified 
whenever the system turns out to be superconducting, since any rotational symmetry breaking pairing would 
be ruled out by the singlet channel; on the other hand, in our study we will only compare the superconducting solution with
a rotationally invariant normal state, and therefore we cannot exclude the possibility that some other ordered phase would
win the competition for the lowest-energy phase.

We can now turn to the problem of establishing the independent slave-boson amplitudes required by our 
model in order to describe a superconducting solution characterized by the $L=S=0$ order parameter
defined in \ceq{eq:sc_order_param}. Using the Wigner-Eckart theorem as for the normal state solution, 
in this case we find the following expression for the
non-zero amplitudes:
\begin{equation} \label{eq:norm_anom_amp}
\begin{array}{l}
 \ds \varphi_{\Gamma n} \;=\; \langle n | \Gamma \rangle \, \Phi (E_\Gamma)  
 + \sum_{q = 1}^3  \left[ \frac{\langle n | \hat{\psi}^{q} | \Gamma \rangle}%
 {\sqrt{\langle \Gamma | (\hat{\psi}^\dagger)^{q} \hat{\psi}^q | \Gamma \rangle}} \Psi(E_\Gamma, 2q) \right.  \\[0.6cm]
 \ds + \left. \frac{\langle n | (\hat{\psi}^\dagger)^{q} | \Gamma \rangle}%
 {\sqrt{\langle \Gamma |  \hat{\psi}^q (\hat{\psi}^\dagger)^{q} | \Gamma \rangle}} \Psi(E_\Gamma, -2q) \right]\!, \qquad
\hat{\psi} =  \sum_a  d^\dagger_{a\uparrow} d^\dagger_{a\downarrow}.
\end{array}
\end{equation}
We have denoted with $\Phi (E_\Gamma)$ the ``normal'' slave-boson amplitudes, which
relate physical and quasi-particle states characterized by the same number of particles, and with
$\Psi(E_\Gamma, \Delta N)$ the ``anomalous'' ones, in which the number of particles characterizing the QP state $|n\rangle$ 
differs by $\Delta N$ from that of the physical state $|\Gamma\rangle$. We remark, however, 
that the presence of non-vanishing anomalous amplitudes is not sufficient, by itself, 
to assure a superconducting solution, the latter requiring the superconducting order parameter $\psi^0_{sc}$,
which is a specific quadratic form in the slave-boson amplitudes, to be finite. 
The normalization factors for the anomalous amplitudes, in \ceq{eq:norm_anom_amp}, are chosen in order to simplify 
the expression for the probability associated with the physical configuration $|\Gamma\rangle$,   
\begin{eqnarray}
P(\Gamma) &=& \sum_n |\varphi_{\Gamma n}|^2 \nn \\ 
&=& |\Phi (E_\Gamma)|^2 + \sum_{\Delta N} |\Psi(E_\Gamma, \Delta N)|^2. \label{eq:probab_dist}
\end{eqnarray}

Using \ceq{eq:norm_anom_amp}, which relates all the slave-boson amplitudes $\varphi_{\Gamma n}$ to
the independent variables $\Phi(E_\Gamma)$ and $\Psi(E_\Gamma,\Delta N)$, we are almost ready 
to write the free-energy functional \eq{eq:free_en} in terms of only $\Phi$'s and $\Psi$'s, obtaining therefore a much smaller
number of saddle-point equations to be solved. The last step needed to achieve this goal, in fact, is to 
evaluate the matrix elements $\langle n |\hat{\mathcal{O}}_d | \Gamma \rangle$ between the eigenstates of $H_\mathrm{loc}$,
which form the basis for the local physical Hilbert space, and the Fock states $|n\rangle_d $ expressed in terms 
of the physical electron operators $d^\dagger_{a\sigma}$.

As discussed in \sec{sec:model}, the effective local Hamiltonian for the electron dynamics of the $\mathrm{{C_{60}}^{n-}}$ ion
is given, in the anti-adiabatic limit, by
\begin{eqnarray}
H_\mathrm{loc} &=& 
 \frac{U}{2} (\hat{n}-3)^2  -\mu \hat{n} \\
 & & +\; J \left[ 2\, {\mathbf{S}} \cdot {\mathbf{S}} + \frac{1}{2} {\mathbf{L}} \cdot {\mathbf{L}} + \frac{5}{6} (\hat{n} -3)^2 \right], \nn
\end{eqnarray} 
where we have included also the chemical potential term, as required by the general formalism described in \sec{sec:formalism},
and we have dropped, for simplicity, the redundant lattice site index. In comparison 
to \ceq{eq:H_loc_JT}, the Coulomb interaction is here written in a particle-hole
symmetric form by properly redefining the chemical potential

We can then readily identify the eigenstates of $H_\mathrm{loc}$ among the atomic multiplets $|\Gamma\rangle$
which are simultaneous eigenstates of the density operator $\hat{n}$ and
of the orbital and spin angular momentum operators ${\mathbf{L}}^2$ and ${\mathbf{S}}^2$, with eigenvalues 
\begin{eqnarray}
E_\Gamma & = & E(n,\ell,s) \label{eq:eigenvalues_nls}  \\
 & = &  \frac{U}{2} (n-3)^2   -\mu n \nn \\
 & &  +\; J \left[ 2 s(s+1) + \um \ell(\ell+1)  + {\ts \frac{5}{6} } (n-3)^2 \right]. \nn
\end{eqnarray}
The degeneracy associated to each eigenvalue is given by
\begin{equation}
g_{[n,\ell,s]} = (2\ell +1)(2s +1)
\end{equation}
and it is therefore natural to choose as a basis set for the corresponding degenerate subspace
the simultaneous eigenstates of both one of the components of ${\mathbf{L}}$ and ${\mathbf{S}}$, say $L_z$\cite{note6} and $S_z$,
so that we can finally set
\begin{equation}
|\Gamma\rangle \equiv \left|\stackrel{\ph{.}}{n},(\ell,\ell_z),(s, s_z)\right\rangle . \label{eq:multiplets}
\end{equation}
These states must then be expressed in terms of the Fock states $|n\rangle_d $, in order to evaluate the matrix elements 
which characterize \ceq{eq:norm_anom_amp}. We note, however, that 
\begin{equation}
 |n\rangle_d \equiv \prod_{a=1}^3 \prod _{\sigma = \uparrow,\downarrow}
 \left(d_{a\sigma}^\dagger\right)^{n_{a\sigma}} \vac \qquad [n_{a\sigma} = 0,1] \label{eq:fock_d}
\end{equation}
are not eigenstates of any of the orbital angular momentum operators $L_a$, making 
the representation of the atomic multiplets \eq{eq:multiplets} in terms of such states a bit involved:
it is more convenient, instead, using the rotational invariance of $H_\mathrm{loc}$, to choose 
an orbital basis for the physical electron operators in which $L_z$ is diagonal,
\begin{eqnarray}
 c^\dagger_{m\sigma} &=& \sum_a U_{ma} d^\dagger_{a\sigma}, \label{eq:orb_rotation} \\
 L_z  &=&  \sum_{m,\,\sigma} m \, c^\dagger_{m\sigma} c_{m\sigma}, \qquad [m = 1,0,-1]    
\end{eqnarray}
so that the corresponding Fock states
\begin{eqnarray}
|n\rangle_c & \equiv & 
 \prod_{m,\,\sigma} \left(c_{m\sigma}^\dagger\right)^{n_{m\sigma}} \vac \\
 &=& \mathcal{U}(U)_{nn'}|n'\rangle_d  \nn
\end{eqnarray}
are eigenstates of both $L_z$ and $S_z$. Since we are assuming the three bands to be degenerate, 
the rotation of the orbital basis \eq{eq:orb_rotation} 
does not change the form of the kinetic term in $H_\mathrm{latt}$, while the expression for the singlet pair-creation operator 
$\hat{\psi} \equiv  \sum_a  d^\dagger_{a\uparrow} d^\dagger_{a\downarrow}$ reads, in the new basis,
\begin{equation}
\hat{\psi} \;=\; c^\dagger_{1\uparrow} c^\dagger_{-1\downarrow} - c^\dagger_{1\downarrow} c^\dagger_{-1\uparrow} - 
 c^\dagger_{0\uparrow} c^\dagger_{0\downarrow}. \label{eq:pair-creation_c}
\end{equation}

\begin{table}[htbp]
\begin{center}
\scriptsize
\begin{tabular}{|c| l c c |c|}
\hline
$\ds \ph{\int} |\Gamma\rangle \ph{\int} $ & $n$ & $(\ell,\,\ell_z)$ & $(s,\,s_z)$ & $\varphi_{\Gamma n}$ \\
\hline
\hline
$|0,0,0\rangle$ & $0$ & $(0,\,0)$ & $(0,\,0)$ & \btc $\Phi(0,0)$\\ $\Psi(0,0;2)$\\ $\Psi(0,0;4)$\\ $\Psi(0,0;6)$ \etc \\
\hline
\hline 
$\left|\uparrow,0,0 \right\rangle$ & $1$ & $(1,\,1)$ & $(\um,\,\um)$ & \btc $\Phi(1,1)$\\ $\Psi(1,1;2)$\\ $\Psi(1,1;4)$ \etc \\
\hline
\hline
$\left|\uparrow \downarrow,0,0\right\rangle$ & $2$ & $(2,\,2)$ & $(0,\,0)$ & \btc $\Phi(2,2)$\\ $\Psi(2,2;2)$ \etc \\
\hline
$\left|\uparrow, \uparrow, 0\right\rangle$ & $2$ & $(1,\,1)$ & $(1,\,1)$ & \btc $\Phi(2,1)$\\ $\Psi(2,1;2)$ \etc \\
\hline
$ \frac{1}{\sqrt{3}} \left[ \left|\uparrow, 0, \downarrow \right\rangle \stackrel{\ph{.}}{-} \left|\downarrow, 0, \uparrow \right\rangle 
     - \left|0, \uparrow \downarrow, 0 \right\rangle \right] $ 
 & $2$ & $(0,\,0)$ & $(0,\,0)$ & \btc $\Phi(2,0)$\\ $\Psi(2,0;-2)$ \\ $\Psi(2,0;2)$ \\$\Psi(2,0;4)$ \etc \\
\hline
\hline
$\ph{\int^\int_|}$ $\left|\uparrow \downarrow, \uparrow,0\right\rangle$ $\ph{\int^\int_|}$ & $3$ & $(2,\,2)$ & $(\um,\,\um)$ & $\Phi(3,2)$ \\
\hline
$ \frac{1}{\sqrt{2}} \left[ \left|\uparrow, \uparrow \downarrow, 0 \right\rangle \stackrel{\ph{.}}{+} \left|\uparrow \downarrow, 0, \uparrow \right\rangle \right] $  & $3$ & $(1,\,1)$ & $(\um,\,\um)$ & \btc $\Phi(3,1)$\\ $\Psi(3,1;-2)$ \\ $\Psi(3,1;2)$ \etc \\
\hline
$\ph{\int^\int_|}$ $\left|\uparrow, \uparrow, \uparrow \right\rangle$ $\ph{\int^\int_|}$& $3$ & $(0,\,0)$ & $(\frac{3}{2},\,\frac{3}{2})$ & $\Phi(3,0)$ \\
\hline
\hline
$\left|\uparrow \downarrow, \uparrow \downarrow,0\right\rangle$ & $4$ & $(2,\,2)$ & $(0,\,0)$ & \btc $\Phi(4,2)$\\ $\Psi(4,2;-2)$ \etc \\
\hline
$\left|\uparrow \downarrow, \uparrow, \uparrow \right\rangle$ & $4$ & $(1,\,1)$ & $(1,\,1)$ & \btc $\Phi(4,1)$\\ $\Psi(4,1;-2)$ \etc \\
\hline
$\! \frac{1}{\sqrt{3}} \left[ \left| \downarrow, \uparrow \downarrow, \uparrow \right\rangle \stackrel{\ph{.}}{-} \left| \uparrow, \uparrow \downarrow, \downarrow \right\rangle  - \left|\uparrow \downarrow, 0, \uparrow \downarrow \right\rangle \right] \!$ 
  & $4$ & $(0,\,0)$ & $(0,\,0)$ & \btc $\Phi(4,0)$\\ $\Psi(4,0;-4)$ \\ $\Psi(4,0;-2)$ \\$\Psi(4,0;2)$ \etc \\
\hline
\hline
$\left|\uparrow \downarrow, \uparrow \downarrow, \uparrow \right\rangle$ & $5$ & $(1,\,1)$ & $(\um,\,\um)$ & \btc $\Phi(5,1)$\\ $\Psi(5,1;-4)$\\ $\Psi(5,1;-2)$ \etc \\
\hline
\hline
$\left|\uparrow \downarrow,\uparrow \downarrow,\uparrow \downarrow \right\rangle$ 
  & $6$ & $(0,\,0)$ & $(0,\,0)$ & \btc $\Phi(6,0)$\\ $\Psi(6,0;-6)$\\ $\Psi(6,0;-4)$\\ $\Psi(6,0;-2)$ \etc \\
\hline
\end{tabular}
\end{center}

\caption{Electronic eigenstates of the $\mathrm{{C_{60}}^{n-}}$ molecular ion, and the
corresponding slave-boson amplitudes: the latter are selected in order
to preserve, at the saddle-point level, the rotational invariance of the local Hamiltonian.}\label{tab:amplitudes}
\end{table}

The representation of the atomic eigenstates $|\Gamma\rangle$ in terms of the Fock states $|n\rangle_c$ is listed
in Table \ref{tab:amplitudes}, where they are classified according to the quantum numbers $(n,\ell,s)$ which 
determine, through \ceq{eq:eigenvalues_nls}, 
the corresponding eigenvalues; note, however, that for a given value of the particle number $n$,
the Pauli-principle prevents the orbital and spin angular momenta
$\ell$ and $s$ to take independent values, so that each degenerate-multiplet can actually be identified specifying only two quantum numbers,
$n$ and $\ell$. For each multiplet, we have written out explicitly only the component characterized by the maximum value of $\ell_z$ and $s_z$,
all the other components being easily obtainable from the former by repeatedly acting on it with the lowering 
operators
\begin{eqnarray}
 L_{-} &=& \sqrt{2} \sum_\sigma \left( c^\dagger_{0 \sigma} c_{1 \sigma}^{\ph{\dagger}} + c^\dagger_{-1 \sigma} c_{0 \sigma}^{\ph{\dagger}} \right), \\
 S_{-} &=& \sum_m c^\dagger_{m \downarrow} c_{m \uparrow}.
\end{eqnarray}   
The last column of the Table contains the independent slave-boson amplitudes $\Phi(n,\ell)$ and $\Psi(n,\ell;\Delta N)$ 
associated, according to \ceq{eq:norm_anom_amp}, to all the components of a given degenerate-multiplet: the total
number of amplitudes required by our model is therefore 35, if we consider the full rotationally-invariant solution, 
while it reduces to 13 if we force the system into the normal state, setting $\Psi(n,\ell;\Delta N) \equiv 0$.

\subsubsection*{Lagrange multipliers}

The Lagrange multipliers $\A^0$, $\Lambda_{\alpha \beta}$ and $\Pi_{\alpha \beta}$, 
introduced in Secs.~\ref{sec:gauge} and \ref{sec:saddle-point} to enforce the constraint equations \eqs{eq:const1}{eq:const3}, 
form, together with the slave-boson amplitudes, 
the set of variables on which the free-energy functional \eq{eq:free_en} 
is defined. However, similarly to what we established in the case of
the slave-boson amplitudes, we must note that the symmetries of our problem greatly reduce the 
number of independent Lagrange multipliers required for the solution of the model,
and in the following we will identify the form of such variables.

Denoting with $\alpha \equiv (m,\sigma)$ both the orbital and spin indices, 
the constraints \eqs{eq:const1}{eq:const3} read, at the saddle-point level,
\begin{eqnarray}
 \sum_{\Gamma} P(\Gamma) &=&1 ,\\
 \mathcal{Q}^N_{\alpha \beta} &=& \langle f_{\alpha}^\dagger f_{\beta} \rangle \nn \\
 &=& \sum_\kk G_f(\kk,0^-)_{\beta \alpha}, \label{eq:QP_const_N}\\
 \mathcal{Q}^A_{\alpha \beta} &=& \langle f_{\alpha}^\dagger f_{\beta}^\dagger \rangle \nn \\
 &=& \sum_\kk F_f(\kk,0^-)_{\beta \alpha}, \label{eq:QP_const_A}
\end{eqnarray}
where $P(\Gamma)$ is the probability distribution defined in \ceq{eq:probab_dist},
$\mathbf{G}_f(\kk, \tau)$ and $\mathbf{F}_f(\kk, \tau)$ are the normal and anomalous 
quasiparticle Green's functions, and
\begin{eqnarray}
\mathcal{Q}^N_{\alpha \beta} &\equiv &
   \sum_{\Gamma nn'} \varphi_{\Gamma n}^* \varphi_{\Gamma n'} \langle n'| f_{\alpha}^\dagger f_{\beta} |n \rangle, \\
\mathcal{Q}^A_{\alpha \beta} &\equiv &   
   \sum_{\Gamma nn'} \varphi_{\Gamma n}^* \varphi_{\Gamma n'} \langle n'| f_{\alpha}^\dagger f_{\beta}^\dagger |n \rangle
\end{eqnarray}
are defined as the normal and anomalous quasiparticle density matrices. As for $P(\Gamma)$, we can 
then make use of \ceq{eq:norm_anom_amp} in order to rewrite the left-hand side of Eqs.~\eq{eq:QP_const_N} and \eq{eq:QP_const_A}
directly in terms of the independent slave-boson amplitudes, obtaining the following expressions: 
\begin{eqnarray}
\mathcal{Q}^N_{\alpha \beta} & = & \mathcal{Q}^N[\Phi,\Psi] \, \delta_{\alpha \beta} , \\
\mathcal{Q}^A_{\alpha \beta} & = & \mathcal{Q}^A[\Phi,\Psi] \, \delta_{\alpha,\bar{\beta}} \, (-1)^{\eta_\alpha}, \\
    \eta_\alpha & = & m + \sigma +\um, \qquad [\sigma = \pm \um] \nn
\end{eqnarray}
where the spin and orbital indices of $\bar{\alpha}$ are opposite to those of $\alpha$. The
specific choice we have made for the independent slave-boson amplitudes thus reflects in a very simplified
form of the quasiparticle density matrices, and it is not hard to recognize in this
structure the symmetry properties which characterize our model,
i.e., the band-degeneracy and the $SU(2)\otimes O(3)$ rotational invariance. 
The quasiparticle energy matrix $\mathbf{h}(\kk)$\cite{note7}
must then be rotationally invariant as well, 
in order to yield quasiparticle expectation values with the same structure of the QP density matrices:
$$
\langle f_{\alpha}^\dagger f_{\beta} \rangle \propto \delta_{\alpha \beta}, \qquad 
\langle f_{\alpha}^\dagger f_{\beta}^\dagger \rangle \propto \delta_{\alpha,\bar{\beta}} \, (-1)^{\eta_\alpha}.
$$
The kinetic part of $\mathbf{h}(\kk)$, namely $\RR^\dagger \tilde{\ee}(\kk) \RR$, 
is guaranteed to be $SU(2)\otimes O(3)$ invariant, since it depends only on the degenerate band dispersion 
$\epsilon_{\alpha\beta}(\kk) = \epsilon(\kk) \delta_{\alpha \beta}$ and
on the slave-boson amplitudes $\varphi_{\Gamma n}[\Phi, \Psi]$ (through the $R$-matrices) 
which have been properly selected in order to yield rotationally invariant solutions. 
On the other hand, the Lagrange multipliers matrices $\mathbf{\Lambda}$ and $\mathbf{\Pi}$
are, in principle, two generic Hermitian and antisymmetric matrices, respectively, and 
we must therefore set 
\begin{equation}
\Lambda_{\alpha \beta} = \Lambda \, \delta_{\alpha \beta}, \qquad 
\Pi_{\alpha \beta} = \Pi \, \delta_{\alpha,\bar{\beta}} \, (-1)^{\eta_\alpha},
\end{equation}
in order to guarantee the rotational invariance of the quasiparticle Hamiltonian. 
   
In the end, we are left with just three Lagrange multipliers, $\A^0$, $\Lambda$, and $\Pi$,
to which we can eventually add the chemical potential $\mu$ if we decide to solve the model
keeping the physical electron density fixed: in the latter case, in fact, the chemical potential 
plays the role of a Lagrange multiplier for the number equation
\begin{equation}
 n_\mathrm{phys} \equiv \sum_{m,\,\sigma} \langle c^\dagger_{m\sigma} c_{m\sigma} \rangle = \sum_\Gamma n(\Gamma) P(\Gamma)
\end{equation}
rather than being an external parameter as in the grand-canonical ensemble.

\subsubsection*{Spectral weight and low-energy excitations}

As shown in \sec{sec:Green_fun}, the expression for the
quasiparticle spectral weight matrix $\mathbf{Z}$, defined as 
\begin{equation}
\mathbf{Z} = \left[ \ident - \frac{\de \mathbf{\Sigma}}{\de \omega} \right]^{-1}_{\omega = 0},
\end{equation}
is given, in terms of the slave-boson amplitudes, by
\begin{equation}
\mathbf{Z} = \RR\RR^\dagger,
\end{equation}
$\RR[\{\varphi\}]$ being the matrix which relates the physical electron operators 
to the quasiparticle ones (see Eqs. (\ref{eq:R-matrices}) and (\ref{eq:RR-matrix})
for its definition). In our model, this relation reads
\begin{equation}
 c_{m\sigma}^\dagger \,=\, r_p[\Phi, \Psi]^* f_{m\sigma}^\dagger
 + (-1)^{m + \sigma +\um} \, r_h[\Phi, \Psi] \, f_{-m \bar{\sigma}}^{\ph{\dagger}}, 
\end{equation}
and it can be easily recognized, as in the case of the quasiparticle density matrices,
the specific structure of the normal ($r_p$) and anomalous ($r_h$)
terms, dictated by the symmetries of the problem. Inserting this relation
in the definition of the quasiparticle weight matrix, we finally obtain
\begin{equation}
\mathbf{Z} = \left( |r_p|^2 + |r_h|^2 \right) \ident,
\end{equation}
which states that, for rotationally invariant solutions to this model, all the electronic degrees of freedom are renormalized
by the same factor and do not mix each other due to the interaction terms.

Besides the quasiparticle spectral weight, we can actually extract, from the 
saddle-point values of the slave-boson fields and Lagrange multipliers, 
the entire low-energy spectrum of the system, i.e., its coherent single-particle 
excitations.
They are defined as the frequency-poles of the physical electron propagator 
\begin{equation}
\mathbf{D}_c(\kk,\omega) \,=\, \RR \, \mathbf{D}_f(\kk,\omega) \RR^\dagger \,=\, \RR \, [\omega - \mathbf{h}(\kk)]^{-1} \RR^\dagger
\end{equation}
and, in terms of the saddle-point variables, they are given by the six-fold degenerate branches
\begin{eqnarray}
E_\mathrm{1p}(\kk) &=& \pm 
   \sqrt{ \ds \left[ \epsilon_\kk \left( |r_p|^2 + |r_h|^2 \right) + \lambda \right]^2 + |\tilde{\Delta}|^2 }, \ph{\int_|} \label{eq:1p_eigenval} \\
 \lambda &=& \frac{\Lambda \left( |r_p|^2 - |r_h|^2 \right) + r_p r_h \Pi + (r_p r_h \Pi)^* }{|r_p|^2 + |r_h|^2}, \nn \\
 \tilde{\Delta} &=& \frac{ r_p^2 \,\Pi - (r_h^2 \,\Pi)^* - 2 \Lambda r_p r_h^*}{|r_p|^2 + |r_h|^2}. \nn
\end{eqnarray}
From \ceq{eq:1p_eigenval} we can then readily establish the expression for the low-energy spectral gap, 
\begin{equation}
\begin{array}{c}
 \ds \Delta \,=\, \sqrt{\ds |\tilde{\Delta}|^2 + R_2\! \left[ |\lambda| - {\ts\frac{W}{2}} \left( |r_p|^2 + |r_h|^2 \right) \right]}, \\[0.4cm]
 R_2(x) \equiv x^2 \, \theta(x),
\end{array}
\end{equation}
where we have assumed, for the free-electron dispersion, $\epsilon_\kk \in \left[-\frac{W}{2}, \frac{W}{2}\right]$, 
$W$ being the uncorrelated bandwidth. It is important, however, to keep in mind that
the onset of superconductivity in the system is signaled by the presence of a non-zero order parameter $\psi^0_{sc}$
rather than by the opening of a gap in the spectrum: these
two quantities, in fact, are directly proportional to each other only in the weak-coupling regime ($J/W \ll 1$ and $ U \lesssim J$),
where we find the solution to be BCS-like, becoming instead disentangled for stronger values of either the electron correlation $U$
or the pairing attraction $J$.

\section{Illustration of results}\label{sec:results}
 
As mentioned previously, the major strength of slave-boson approaches relies
in the possibility of obtaining, with a relatively low computational effort, 
approximate analytical solutions which describe quite well, on a qualitative footing, the effects of electronic 
interactions on the low-energy part of the spectrum. With such methods we can thus investigate the behavior of 
a given model over the entire range of variability of the parameters on which the model is defined.

Using the slave-boson representation introduced in the 
previous section for the description of superconducting fullerides, 
we will here illustrate the behavior of the saddle-point solutions across the 
zero-temperature ``phase diagram'', where the external parameters of the model are represented 
by the electron density $n_\mathrm{phys}$ and the two ratios $U/W$ and $J/W$, which measure, respectively,
the strength of the Coulomb and Jahn-Teller interactions with respect to the kinetic bandwidth $W \propto t$.
We will primarily focus on the half-filled case, where it can be found the most interesting
experimental feature of these systems, namely the relatively high superconducting critical temperature in comparison to
the strength of the pairing coupling, and then we will briefly analyze how the superconducting behavior
extends to finite values of doping. The solutions are obtained using, for simplicity, a flat density of states, $D(\epsilon) = \frac{1}{W}$, 
since the qualitative behavior of the system does not depend much on the specific form of $D(\epsilon)$.

\subsection{Half-filling ($n_\mathrm{phys} = 3$)}

\begin{figure}[t]
\begin{center}
\includegraphics[width=7.5cm]{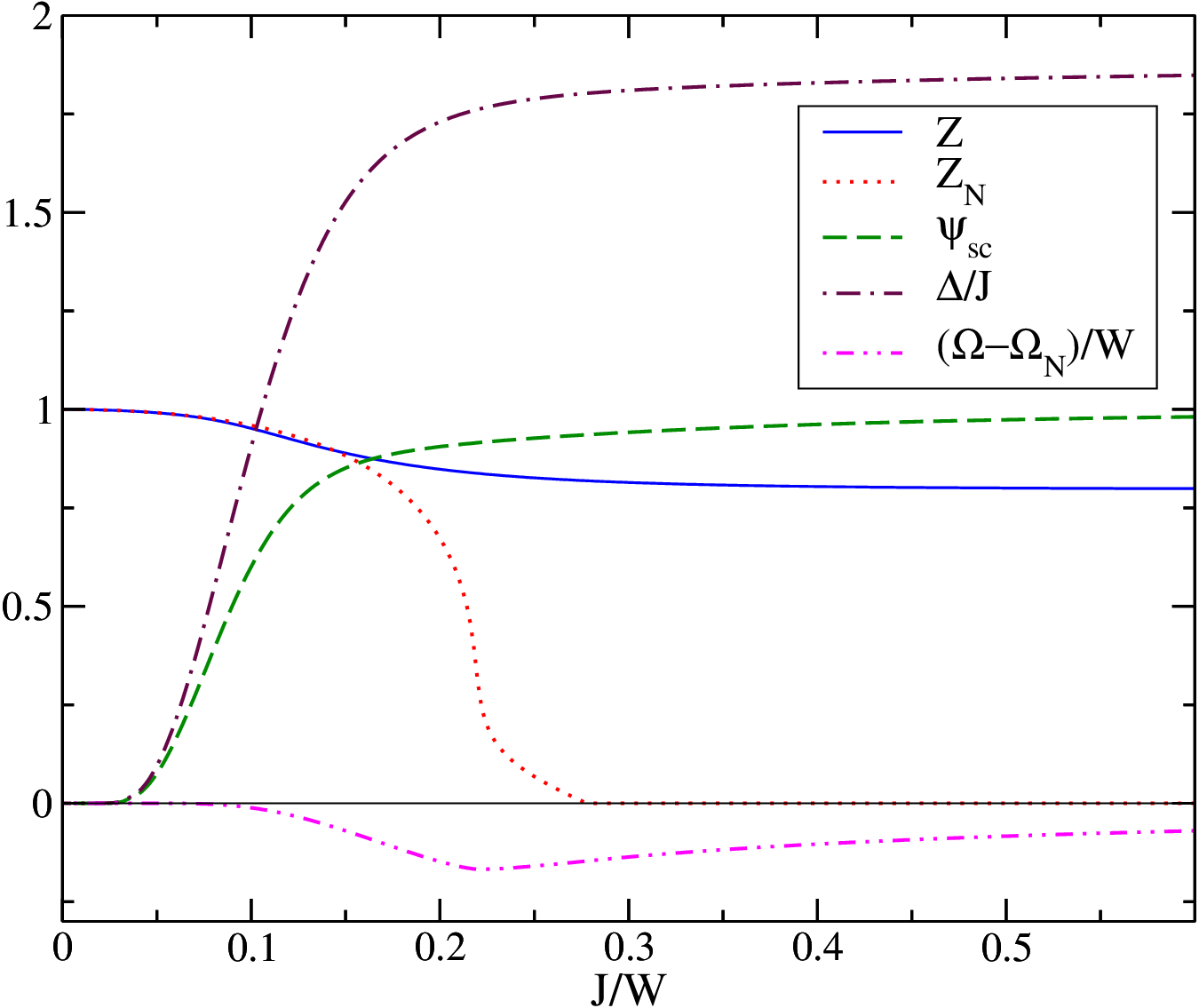}
\end{center}
\caption{(Color online) Normal and superconducting solutions at half-filling and $U=0$: $Z_N$ and $Z$ are the corresponding quasiparticle weights, 
$\Delta$ and $\psi_{sc}$ are the superconducting gap and order parameter,  
$(\Omega - \Omega_N)$ represents the free-energy difference between the normal and the superconducting state.}\label{fig:n3_U0}
\end{figure}

We begin the analysis of the half-filled model illustrating, in \fig{fig:n3_U0}, the normal and superconducting 
solutions obtained, at $U=0$, for increasing values of the Jahn-Teller coupling $J/W$. Since we are turning off the Coulomb repulsion, this is a purely attractive model, with
the Jahn-Teller coupling playing the role of an attractive local interaction acting on spin and orbital degrees of freedom;
the physics of this system is therefore characterized by the competition between singlet formation and kinetic delocalization,
and we find the results of this competition to be remarkably different whether we are considering a purely normal-state solution
or we are allowing for a superconducting order parameter. As expected for a purely attractive interaction, the superconducting solution is always energetically favored at finite $J$.

As soon as the pairing interaction $J$ is turned on, the behavior of the normal-state solution 
is initially characterized by a slow decrease of the quasiparticle weight $Z_N$ 
from the non-interacting value $Z_N(0)=Z(0)=1$, which is then followed by a steep
descent towards zero for $J/W \gtrsim 0.2$; finally, when the coupling is further increased, the metallic state turns into 
an insulating one, where all the electrons are locked in local singlets
formed by two or four electrons\cite{note9}, 
the binding energy of the singlet configuration 
being much more favorable than the kinetic energy gain associated to the electron hopping. 
This state is analogous to the paired insulator found, at strong coupling, in the normal solution of the attractive Hubbard model 
within DMFT\cite{PairedInsulator}. 
\begin{figure}[b]
\begin{center}
\hspace{-0.4cm}
\includegraphics[width=7.8cm]{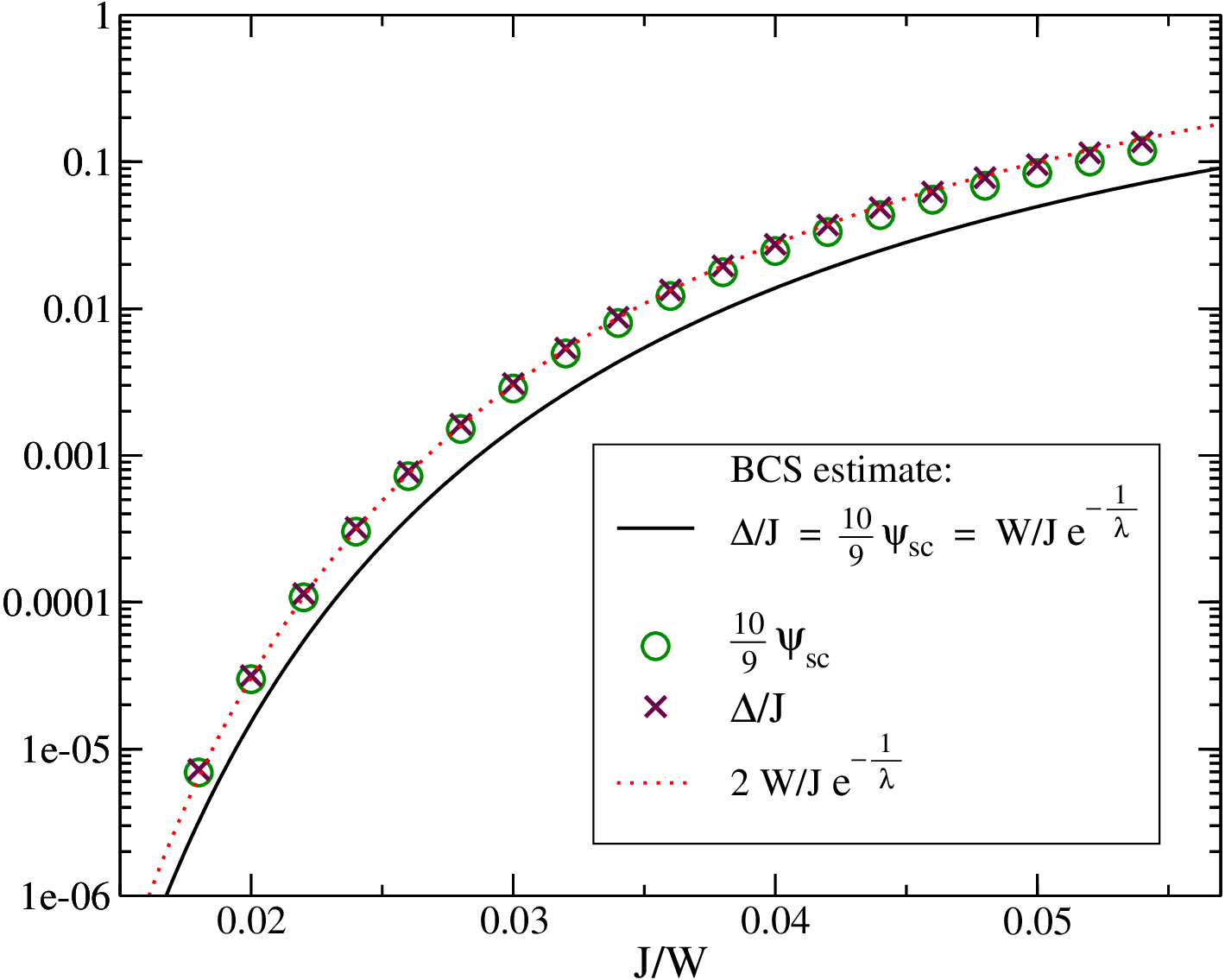}
\end{center}
\caption{(Color online) Superconductivity in the weak-coupling regime: slave-boson solution (circles and crosses) vs.\ BCS estimate.
For $J/W \ll 1$ and $U=0$ the superconducting parameters satisfy the BCS relation $\Delta/J  =  \frac{10}{9} \psi^0_{sc}$
and, apart from a constant prefactor, they follow the BCS functional form $\Delta(J)_{\sss \mathrm{BCS}} = W e^{-\frac{1}{\lambda}}$ 
($\lambda = \frac{10}{3} (J/W)$ is the dimensionless superconducting coupling constant).}\label{fig:n3_U0_BCS}
\end{figure}
On the other hand, if we do allow for superconducting ordering, we find the static singlet-formation mechanism characterizing the
normal solution to be replaced by the more favorable Cooper pair formation, in which the singlet pairs
can still gain some kinetic energy through their propagation: the solution, in this case, is therefore characterized
by a finite quasiparticle weight $Z$ over the entire range of the pairing interaction. We must however notice
that the difference in the behavior of the normal solution between the weak and strong coupling regimes (metallic vs.\
insulating) can still be traced in the behavior of the superconducting solution. In fact, 
for increasing values of $J/W$, we observe a crossover between a weak-coupling BCS-like superconductivity, 
where the gap $\Delta/J$ and the superconducting order parameter $\psi^0_{sc}$ are 
proportional to each other and exponentially small in the pairing coupling (\fig{fig:n3_U0_BCS}), 
\begin{equation}
\ts \Delta/J \; = \; \frac{10}{9} \psi^0_{sc} \; \sim \; \frac{1}{\lambda}  e^{-\frac{1}{\lambda}}, 
\qquad  \lambda = \frac{10}{3} (J/W)  \label{eq:BCS_gap_ordparam}
\end{equation}
and a strong-coupling superconductivity associated to Bose-Einstein Condensation (BEC) of preformed pairs, 
where both the gap and the superconducting order parameter
are saturated\cite{BCSBEC}. 
While in the former regime the formation of Cooper pairs subtracts some kinetic energy from
the normal state in order to gain the binding energy associated to the superconducting singlets, 
as evidenced by the lower spectral weight $Z<Z_N$, which corresponds to more localized particles, 
in the large $J/W$ regime, where the local singlets are already formed, 
the energy gain of the superconducting state is due to the finite kinetic energy of
the Cooper pairs in comparison to the static singlets characterizing the insulating normal state\cite{EnergeticBalance}.

\begin{figure}[b]
\begin{center}
\includegraphics[width=7.5cm]{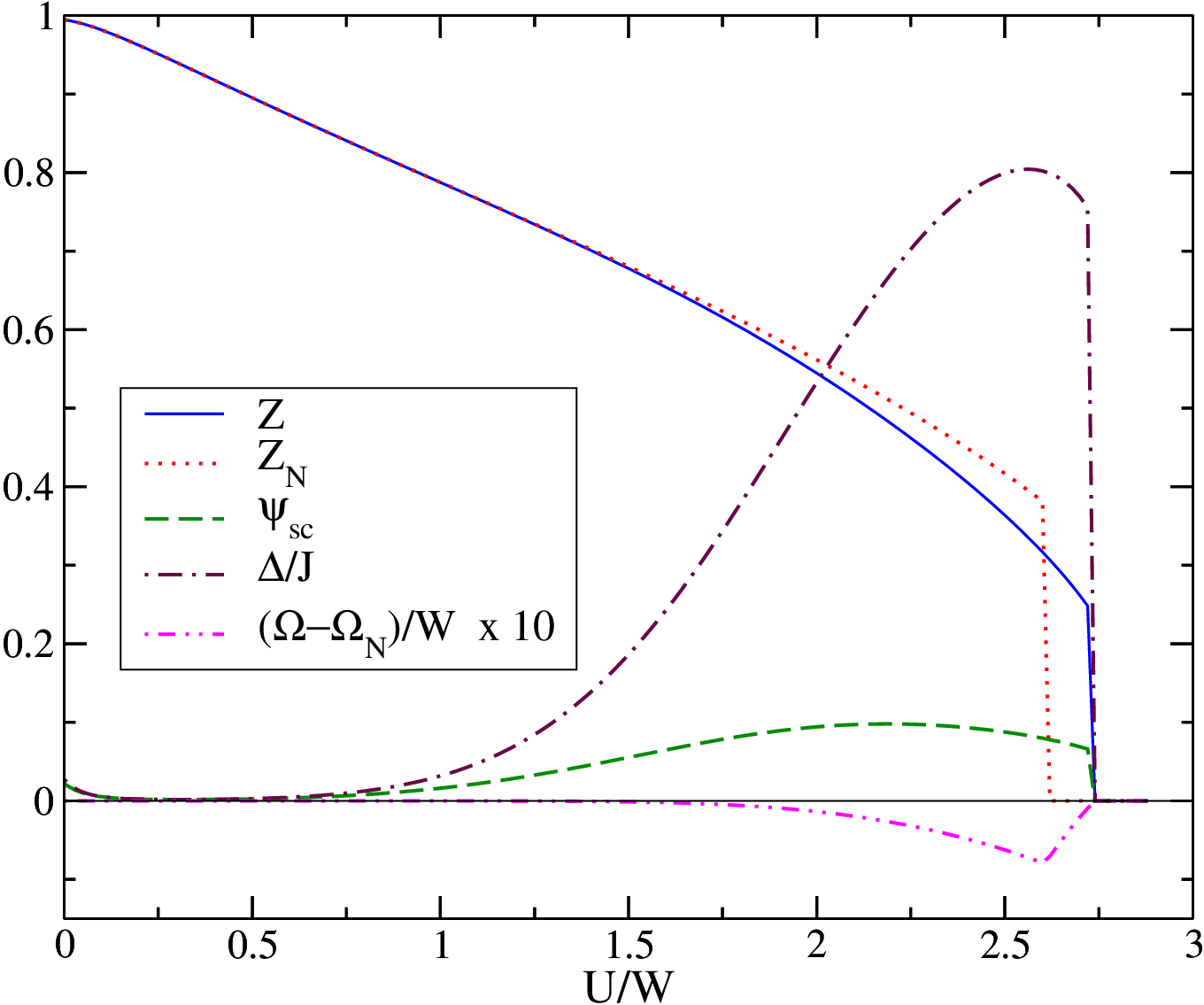}
\end{center}
\caption{(Color online) $U$-dependence of the half-filling solutions at $J/W=0.04$. The 
condensation energy $(\Omega - \Omega_N)/W$ is multiplied by a factor of 10 for visibility reasons.}\label{fig:n3_J0_04}
\end{figure}

The most interesting aspect of the half-filling solutions, however, is represented by 
the behavior of the quasiparticle weight and of the superconducting observables (spectral gap and order parameter)
as functions of the electron correlation $U$, at different values of the Jahn-Teller coupling $J$.

In \fig{fig:n3_J0_04} we plot the $U$-dependence of the normal quasiparticle weight $Z_N$ and of the observables characterizing 
the complete solution ($Z$, $\Delta$ and $\psi^0_{sc}$) at $J/W=0.04$, corresponding to a coupling
strength located in the upper end of the weak-coupling regime, or, in other words, just before the $U=0$ crossover region.
The relevant feature to be noticed in this figure is the non-monotonic behavior of the superconducting parameters 
for increasing values of the electron correlation: while at small $U$ the net effect of the Coulomb repulsion is
to rapidly destroy the superconducting order, as expected in a weak-coupling BCS superconductor (notice the small value of the superconducting order parameter at $U=0$ and its sudden disappearance as soon as $U$ is turned on), at larger $U$ values 
the system turns back superconducting, with an enhancement in the values of $\Delta$ and $\psi^0_{sc}$ in comparison 
to $U=0$, until it undergoes a first-order Mott transition at $U=U_c$, just above the corresponding Mott transition
of the normal state. It is evident the huge enhancement of the superconducting amplitude $\psi^0_{sc}$ with respect to the non-correlated regime.

\begin{figure}[b]
\begin{center}
\includegraphics[width=7.5cm]{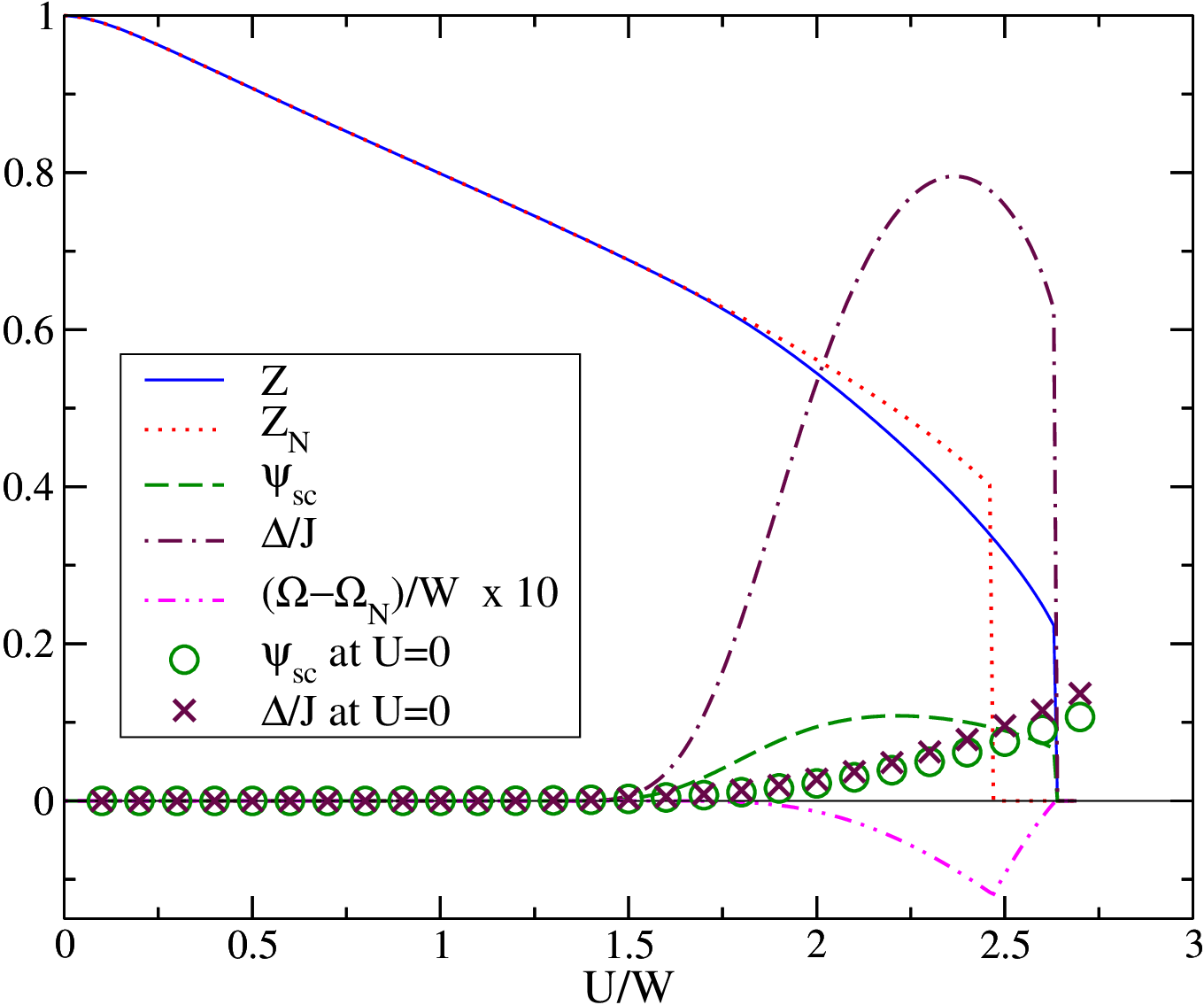}
\vspace{0.4cm}

\includegraphics[width=7.5cm]{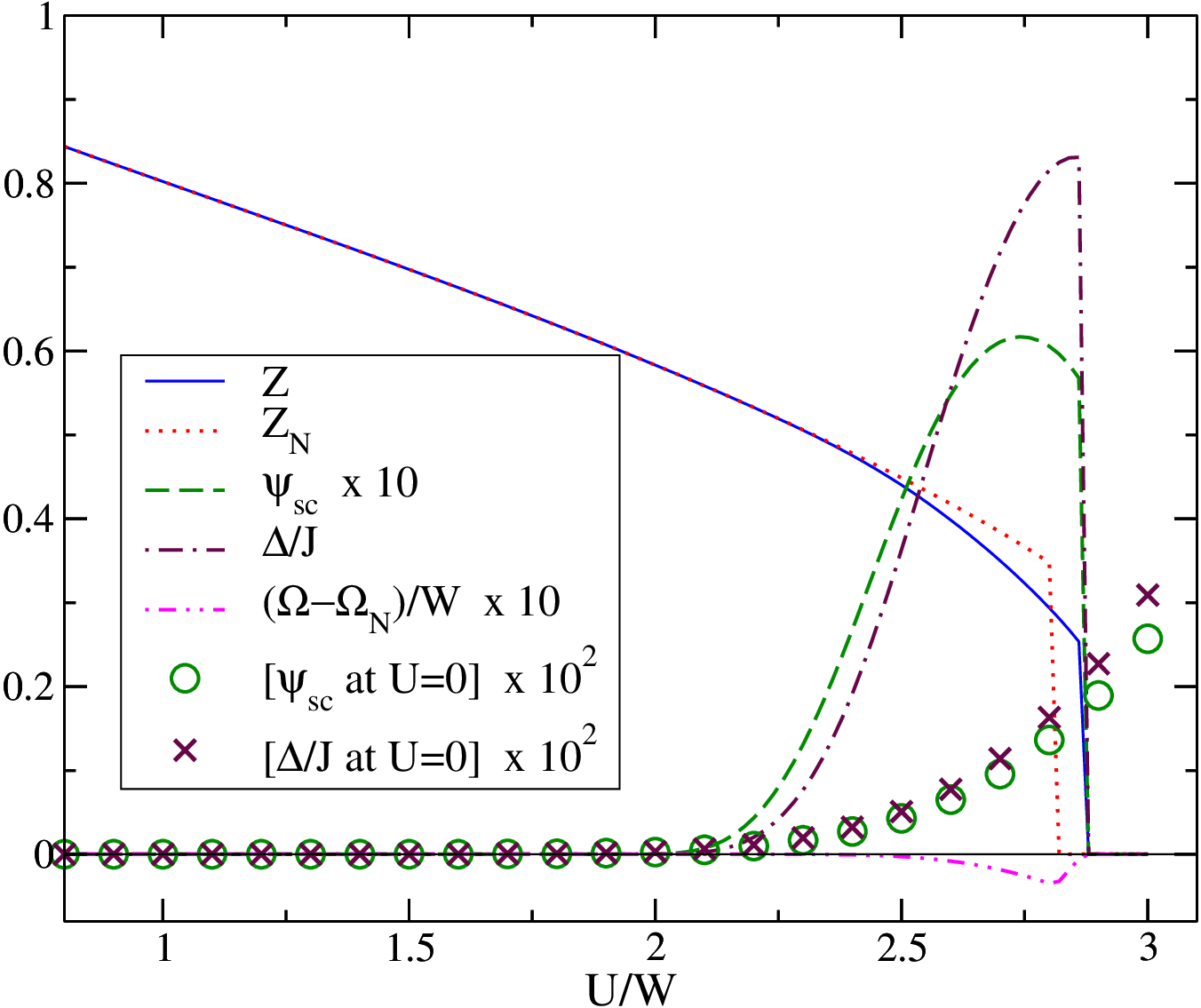}
\end{center}
\caption{(Color online) $U$-dependence of the half-filling solutions for fixed ratios $J/U=$ 0.02 (top) and 0.01 (bottom).
Circles and crosses  represent the superconducting parameters evaluated at $U=0$ 
for the same values of $J$.}\label{fig:n3_JUratio}
\end{figure}

From a physical point of view, the reemergence of superconductivity at large $U$ has been explained
in terms of the ``strongly correlated superconductivity'' scenario put forward using DMFT to solve the same model\cite{Capone-Science,CaponeRMP} and a related simplified model\cite{Capone:04,Schiro}.
The key mechanism is the different effect of the correlation on the various interaction terms: when strong repulsion freezes charge fluctuations, the resulting strongly correlated quasiparticles experience a strongly reduced repulsion, while the attraction is essentially unscreened. As a result, the net effect is that $J/W \ra J/(ZW)$\cite{Capone-Science}: 
when the electrons become more localized, the relative strength of the Jahn-Teller interaction grows in comparison to the renormalized hopping. The precise nature of the interaction, involving orbital and spin degrees of freedom, is crucial in this effect, and proves the ability of our rotationally invariant slave bosons to properly treat every kind of interaction.
The superconducting behavior in this  region is clearly non-BCS-like, as evidenced by the non proportionality between the gap and the order parameter: 
$\Delta/J$ is indeed much larger (up to ten times) than $\psi^0_{sc}$, and its maximum is located much closer 
to the Mott transition than the order parameter's one. On the other hand, the pairing mechanism cannot be explained
within a purely strong-coupling BEC-like picture, since in this case the pairing singlets are not already ``preformed'' in the normal phase (which is either a correlated metal or an $S=1/2$ Mott insulator) and their fraction is much smaller than in standard
BEC superconductivity. We are rather in the presence of a strongly correlated superconductor, 
in which a small local pairing coupling turns out to be enhanced, rather than suppressed, by the effects of a strong on-site repulsion.

\begin{figure}[t]
\begin{center}
\includegraphics[width=7.5cm]{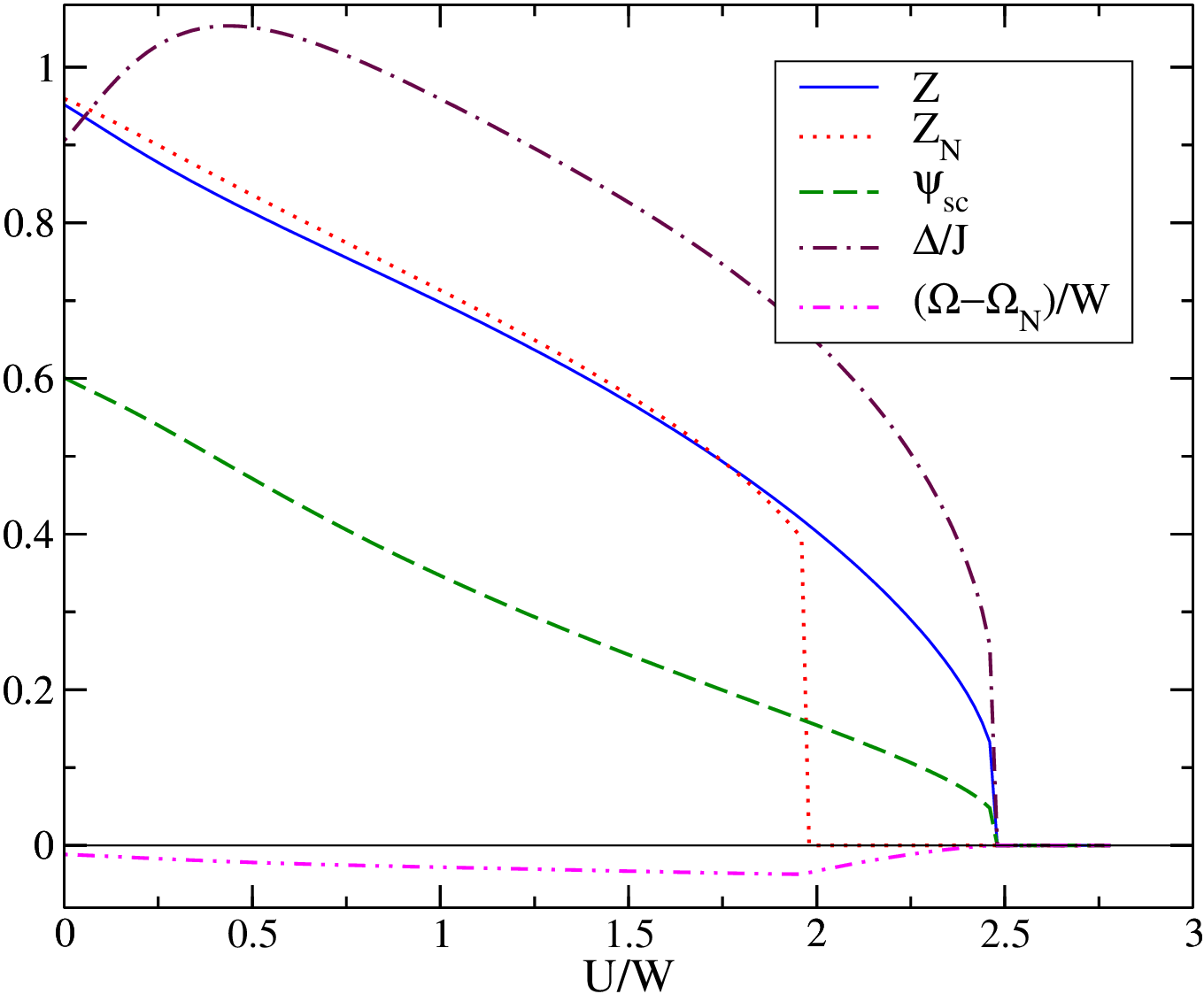}
\vspace{0.4cm}

\includegraphics[width=7.5cm]{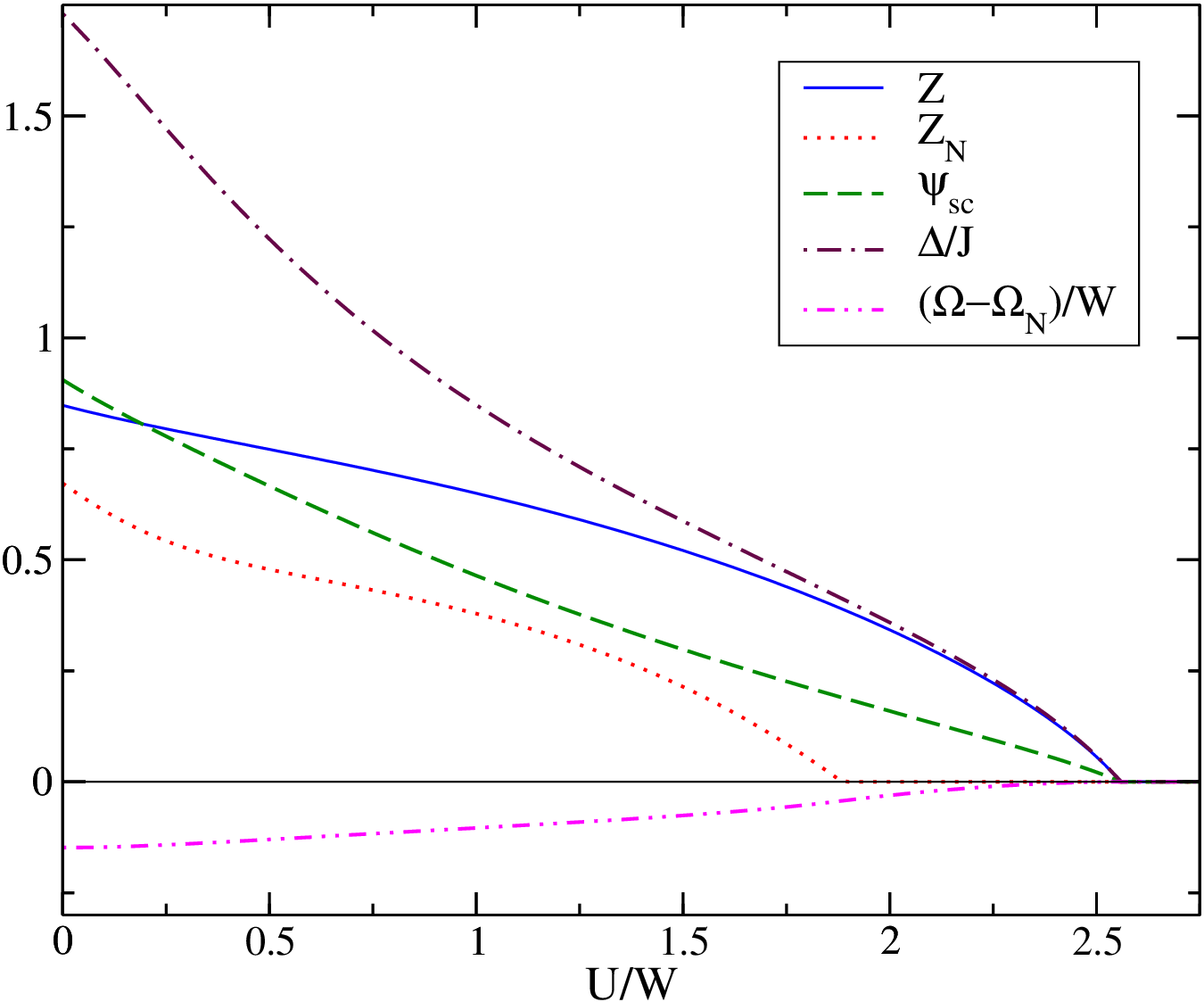}
\end{center}
\caption{(Color online) $U$-dependence of the half-filling solutions at intermediate-to-strong pairing couplings
$J/W=$ 0.1 (top) and 0.2 (bottom).}\label{fig:n3_J0_1}
\end{figure}

The correlation-driven enhancement of superconductivity in the proximity of the Mott transition is even more evident in
\fig{fig:n3_JUratio}, where the solutions are evaluated at a fixed ratio $J/U$ for increasing values of the correlation; together 
with the normal and complete solutions, we have plotted for comparison the (BCS-like) superconducting parameters $\Delta/J$ and $\psi^0_{sc}$
obtained, at $U=0$, for the same values of $J$. Besides the different relation between $\Delta/J$ and $\psi^0_{sc}$
in the correlated case compared to the $U=0$ solutions, these plots emphasize how the enhancement of 
the superconducting gap becomes stronger (up to three orders of 
magnitude in the lower panel) at smaller values of the pairing coupling: for $J/W \ll 1$, indeed, the value
of $\Delta/J$ in the proximity of the Mott transition turns out to be $\ord(1)$, while it is exponentially small in $J/W$
in the BCS regime (see \fig{fig:n3_U0_BCS}).

As already found in DMFT in a two-orbital model\cite{Capone:04}, a completely different scenario is instead observed for larger values of the Jahn-Teller coupling, corresponding to the strong-coupling regime of the $U=0$ attractive model (shown in Fig.~\ref{fig:n3_J0_1}).  
In these cases, in fact, the superconducting order parameter decreases
monotonically with $U$ from the large $U=0$ value, until a weakly first-order Mott transition (which becomes second-order 
when $J/W$ is increased) turns the system into
an insulator; a similar behavior characterizes the superconducting gap, except for an initial rise at small values of $U/W$
in the case $J/W=0.1$. At strong-coupling values of the pairing attraction the superconducting solution
is therefore always energetically-favorable compared to the metallic one, and the electronic correlation has only the effect of
reducing, throughout the non-insulating phase, the superconducting ordering.

\begin{figure}[b]
\begin{center}
\includegraphics[width=7.5cm]{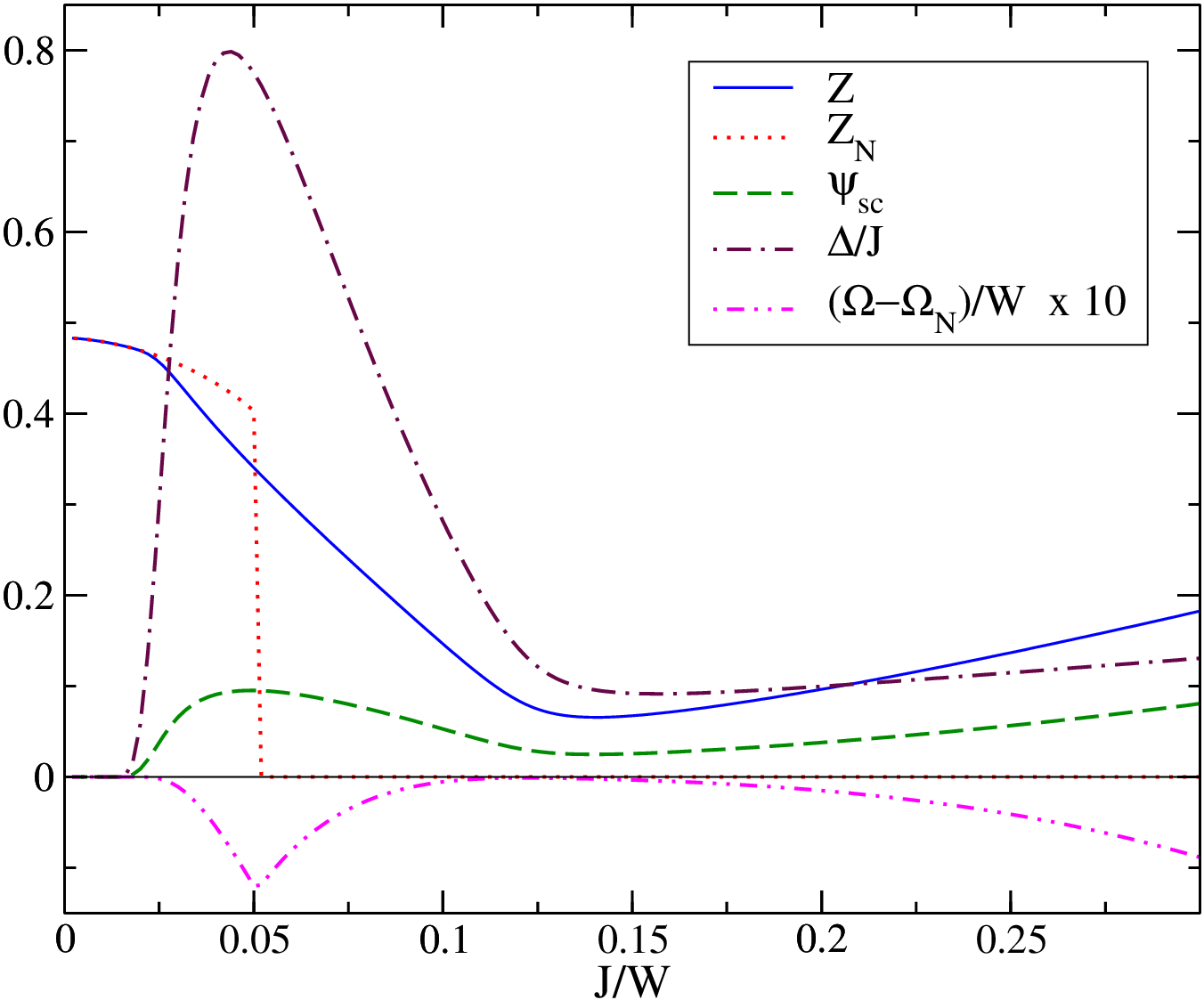}
\vspace{0.4cm}

\includegraphics[width=7.5cm]{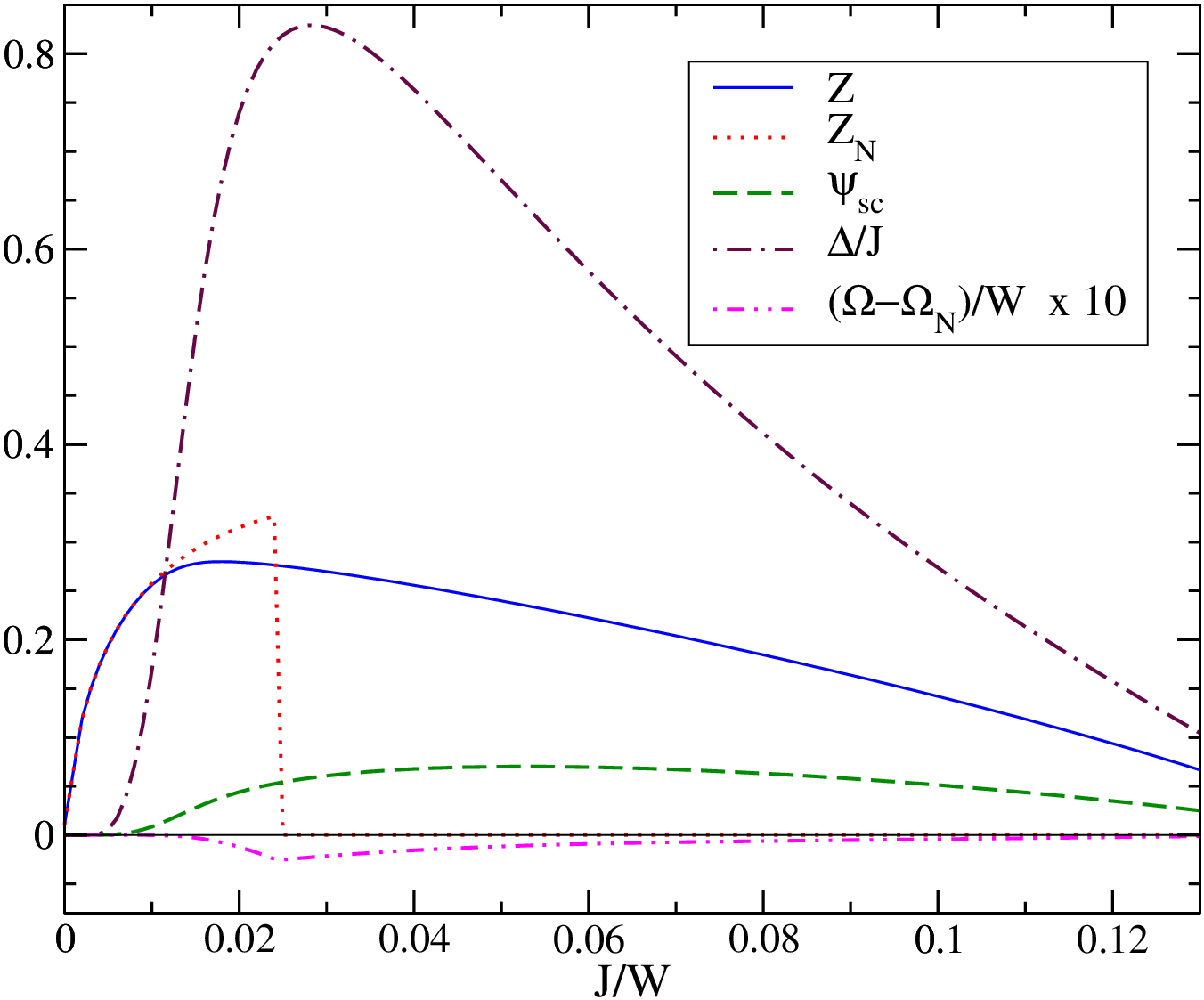}
\end{center}
\caption{(Color online) $J$-dependence of the half-filling solutions in the strongly-correlated regime:
the top panel corresponds to a fixed value of the correlation, $U/W=2.45$,
while in the bottom panel the solution follows the Mott-transition line $U_c(J)$.}\label{fig:n3_U2_45}
\end{figure}

We conclude the analysis of the half-filling solutions showing, in \fig{fig:n3_U2_45}, 
the non-monotonic behaviors of the superconducting parameters
$\Delta/J$ and $\psi^0_{sc}$, as functions of $J$, in the strongly-correlated region of the phase diagram:
in the top panel the value of $U/W$ is held fixed, while in the bottom one it follows the Mott-transition line from below, 
$U(J)=U_c(J)-\delta U$. Combining these results
with the ones discussed previously, we can then infer the existence of a second region in the $J$--$U$ plane, 
beside the strong-coupling BEC-like region at $J \gg U$, in which 
superconductivity is found to be optimal: almost surprisingly, it is located at very
small values of the pairing attraction $J$ and at correspondingly large values of the on-site electron repulsion $U$, just before
the Mott localization transition line.

The results presented in this section confirm that the rotationally invariant slave boson approach is able to accurately
treat interactions of different kinds and, particularly, it is not limited to charge interactions. Indeed we have found that 
the present approach is able to reproduce the relevant physics of a three-orbital Hubbard model for the fullerenes, 
and, in particular, the huge enhancement of phonon-mediated superconductivity in the proximity of the Mott 
transition. The only qualitative aspect of the DMFT solution which is not found in the present study is the second-order
(or very weakly first-order) character of the superconductor-insulator transition.

\subsection{Finite doping ($n_\mathrm{phys} = 3 - \delta$)}

In this section we consider the effect of doping the half-filled three-band model. While this situation can not be experimentally realized, at the moment, in fullerides, the effect of doping is clearly suggestive for analogies with the physics of cuprates.

\begin{figure}[t]
\begin{center}
\hspace{-0.3cm}
\includegraphics[width=7.7cm]{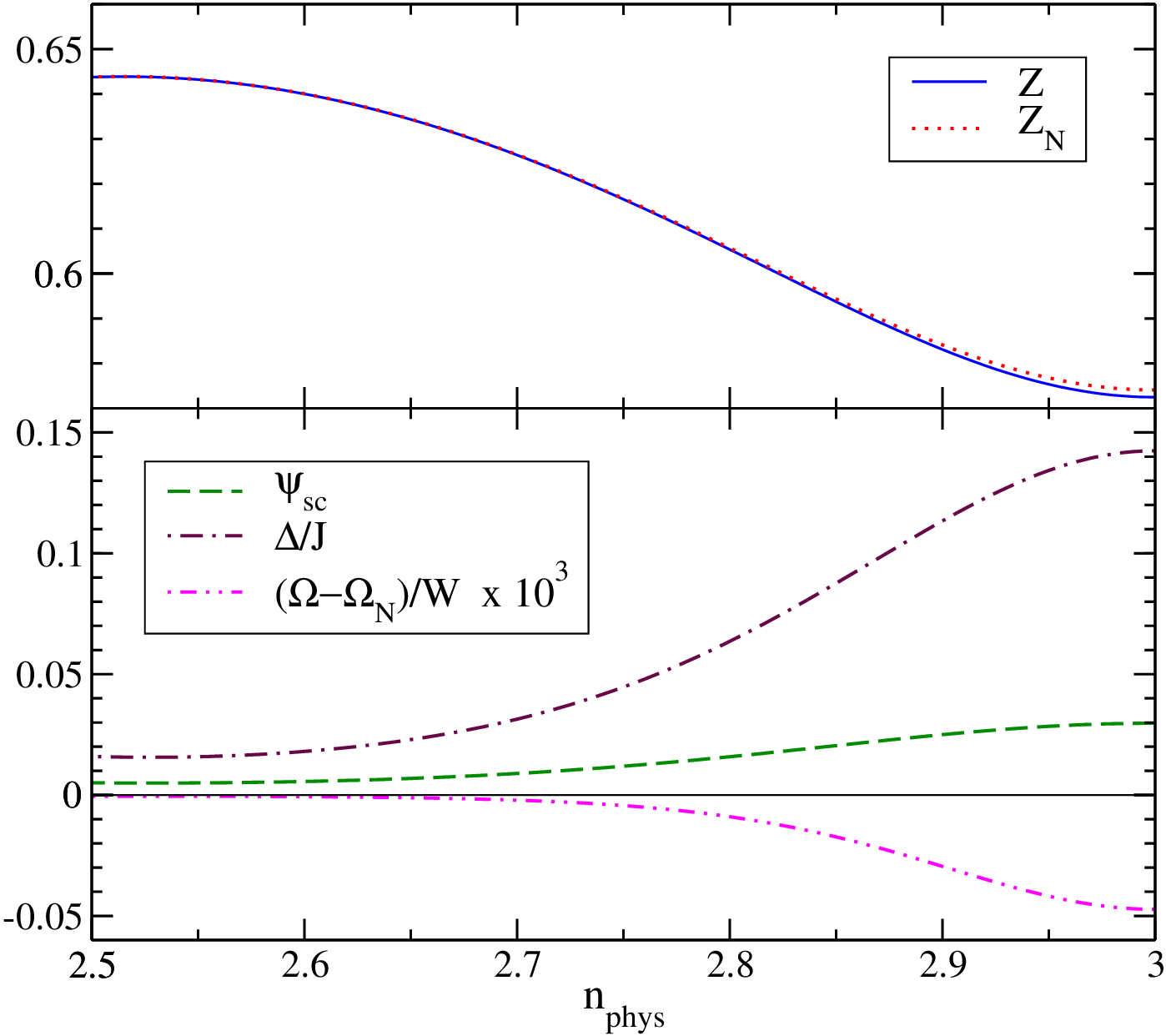}
\vspace{0.5cm}

\includegraphics[width=7.5cm]{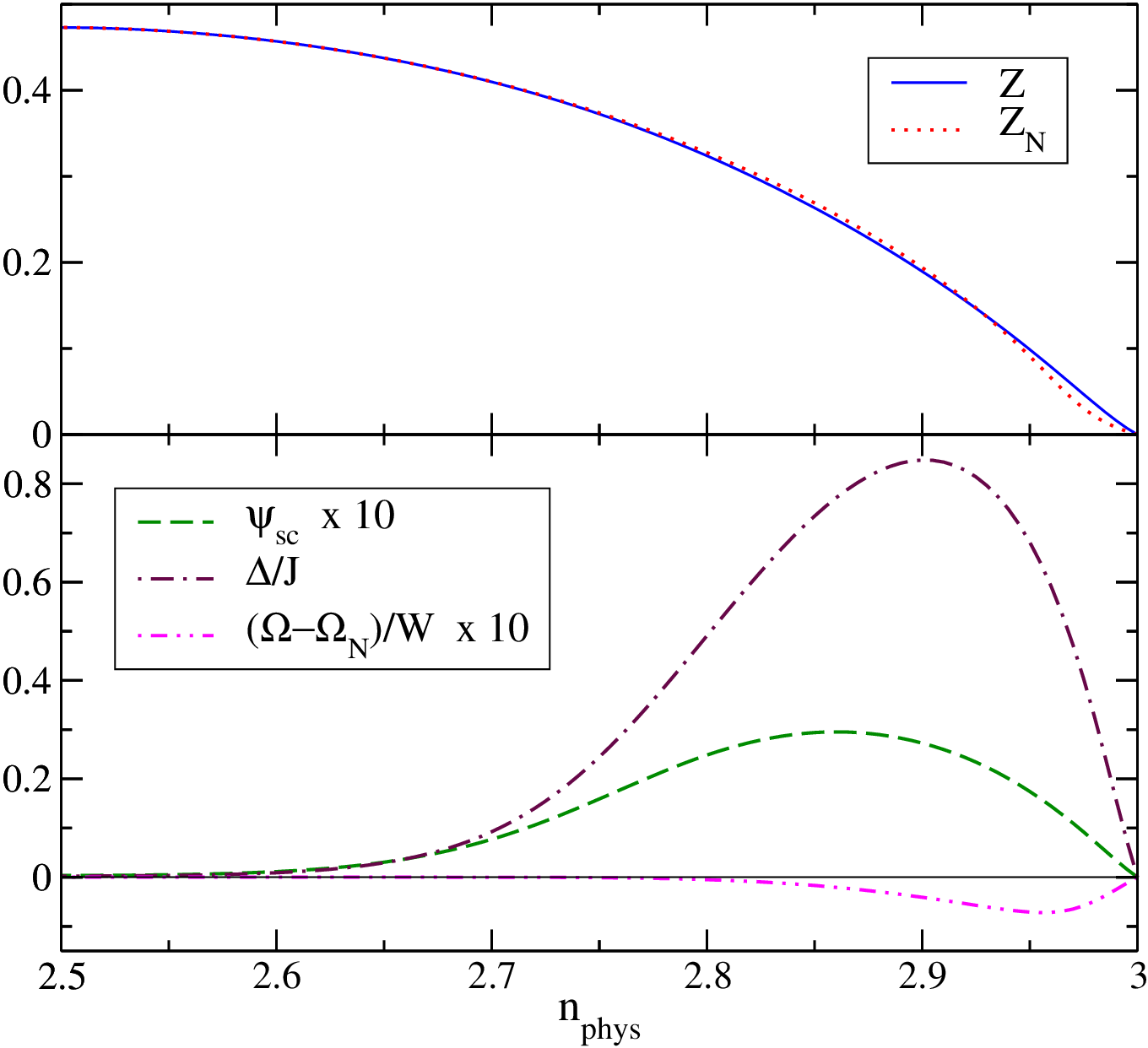}
\end{center}
\caption{(Color online) Doping dependence of the solutions for correlation 
strengths respectively below (top panel, $U/W=2$, $J/W=0.03$) and above (bottom panel, $U/W=5$, $J/W=0.02$) 
the $n_\mathrm{phys}=3$ Mott transition.}\label{fig:doping}
\end{figure}

The behavior as a function of doping, in the neighborhood of $n_\mathrm{phys} = 3$, is 
shown in \fig{fig:doping} for correlation strengths respectively
below and above the critical Mott-transition value $U_c(J)$; in both cases,
however, we have $U \gg J$, so that they both belong to the strongly-correlated region of the phase diagram, 
where the presence of a finite superconducting order parameter is due to the localization-driven enhancement
of the effective pairing coupling.

We find that for $U<U_c$ the superconducting parameters 
decrease monotonically upon doping (eventually vanishing at larger doping values), while the normal and superconducting
quasiparticle weights increase from their finite half-filling values; as long as the superconducting 
order is present, we have $Z<Z_N$. On the other hand, for $U>U_c$ we observe a dome-shaped behavior 
in the superconducting parameters, the gap reaching its maximum at a very small doping value $\delta_\mathrm{opt} \approx 0.1$, while
the order parameter being maximum at a slightly larger optimal doping $\delta^{(\psi)}_\mathrm{opt} \approx 0.15$. 
In this case, both the normal and superconducting quasiparticle weights grow linearly with the doping $\delta$, 
but while in the overdoped region $\delta > \delta_\mathrm{opt}$ we find the standard weak-coupling behavior $Z<Z_N$,
at lower dopings we have $Z>Z_N$.

The behaviors of both the normal and superconducting solutions in the two correlation regions $U \gtrless U_c$
are therefore remarkably different from each other; however, at a closer sight, we find that they can
be actually explained through the same physical mechanism, namely the competition between Mott-localization,
which can eventually enhance the superconducting pairing as we have seen in the discussion of the half-filling solutions, 
and the delocalization tendency introduced by doping. In fact, when the correlation strength at half-filling is not
large enough to completely destroy the quasiparticle coherence (in other words, when the quasiparticle 
degrees of freedom are nor completely frozen), we find the superconducting parameters
to be maximum at zero-doping, where the electrons are more localized and the enhancement 
of the effective pairing coupling is stronger.
When the system is in the Mott-insulating phase, on the contrary, there are no available quasiparticles 
at half-filling in order to develop a superconducting order parameter: the 
reintroduction of quasiparticle coherence due to a finite level of doping becomes then essential in order
to recover a superconducting solution. At small dopings, therefore, the superconducting ordering increases, due 
to the regained coherence of quasiparticles and a still strong enhancement of pairing due to Mott localization; 
for larger values of doping, instead, the loosening of the localization-induced pairing enhancement disfavors the 
superconducting ordering, which turns to decrease as in the $U<U_c$ scenario. It is interesting to 
note that in the underdoped region we have $Z>Z_N$, which means that the formation of superconducting pairs
is energetically more favorable also from the kinetic point of view, compared to the normal state.

\section{Conclusions}\label{sec:conclusions}

We have generalized to superconducting solutions (allowing for the spontaneous breaking of charge symmetry)
the rotationally invariant slave boson approach introduced by Lechermann \textit{et al.}\cite{Lechermann_Georges:07} 
on the basis of the work by Li and W\"olfle\cite{Li_Wolfle:89}.
The crucial ingredient of the rotationally invariant version of slave boson methods is that the boson fields cannot be simply seen as probability amplitudes for the different quasiparticle states, but they are expressed as a non diagonal density matrix that connects the different quasiparticle states to the whole set of physical states. This is easily generalized to include matrix elements between states with different number of particles which allow to describe superconducting ordering.

After a thorough description of the formalism, 
we apply the method to solve a three-band model which has been proposed and studied to understand the properties of alkali-doped fullerides\cite{CaponePRL,Capone:04,CaponeRMP}. This model has been previously solved using DMFT\cite{DMFT} for integer fillings, providing a striking realization of strongly correlated superconductivity, i.e., of a situation in which strong electron-electron correlations favor superconductivity. A crucial element of the model is that the pairing attraction, which can be modelized as an inverted Hund's rule term, only involves spin and orbital degrees of freedom, which are not heavily affected when the charge degrees of freedom are frozen by the proximity to the Mott transition. This interorbital nature of the pairing interaction is crucial to give rise to the correlation-driven enhancement of pairing.
In this light, this model represents an ideal test field for our approach, which is tailored to properly treat interorbital interactions that cannot be expressed in a density-density form.
Indeed the method provides good results for this model, and it is actually the first mean-field-like approach able to reproduce the enhancement of superconductivity observed in DMFT.

The good performance of the method is very encouraging in view of other applications. The most challenging direction is obviously the study of the two-dimensional Hubbard model, which is believed to be the basic model to understand high-temperature superconductivity in the cuprates. While the full solution of the model on a lattice appears too cumbersome, a viable direction is the use of cluster extensions of DMFT, such as the cellular-DMFT\cite{cdmft} or the dynamical cluster approximation\cite{dca}, where the rotationally-invariant slave-boson method can be used as an approximate analytical impurity solver for the cluster Hamiltonian. 
This approach has been used, for example, in Refs.~\onlinecite{Lechermann_Georges:07} and \onlinecite{Ferrero:08} for normal solutions without superconducting ordering. To investigate the superconducting properties of the Hubbard model, on the other hand, our formalism can be applied either to the $2\times 2$ plaquette, where it can be used to better understand the outcomes of fully numerical solutions\cite{plaquette_sc}, or to slightly larger clusters, such as small rectangles in CDMFT or the 5-site ``cross'' in DCA, which have been proposed as ideal compromises between reasonable cluster size and adequate accuracy\cite{Isidori:CDMFT}.

\begin{acknowledgments}

We acknowledge useful discussions with C. Castellani and A. Georges. M. C. acknowledges financial support of MIUR PRIN 2007 Prot. 2007FW3MJX003.

\end{acknowledgments}



\begin{thebibliography}{99}
\bibitem{note3}
The rotation necessarily involves both the physical electron and quasiparticle orbital basis: 
these, in fact, must coincide in the present representation,
in accordance to the matching between physical and QP Fock states imposed by \ceq{eq:density_phys_state}.
\bibitem{noteA0}
To be precise, the transformation law of $\A_i^0$ yields ${\cal L} \ra {\cal L} + i \sum_i\detau \xi_i^0(\tau)$,
which corresponds to a phase shift of the action $S \ra S + i \sum_i [\xi_i^0(\beta) - \xi_i^0(0)]$.
However, the periodic boundary conditions on the boson fields imply that $\xi_i^0(\beta) = \xi_i^0(0) \mod 2\pi$, 
so that $e^{-S}$ remains invariant.
\bibitem{note4} Formally, in the anti-adiabatic limit we have
$E_\mathrm{JT} = \sum_\nu \hbar \omega_\nu g_\nu^2$, with $\omega_\nu$ and $g_\nu$ representing, respectively, the vibron frequencies 
and the electron-vibron couplings which characterize the Jahn-Teller Hamiltonian $H_\mathrm{JT}$.
\bibitem{note6} We can identify $L_a$ as the Cartesian components of ${\mathbf{L}}$ and set, by convention, $L_3 \equiv L_z$.
\bibitem{note7} See \ceq{eq:QP_h_matrix}
for the definition of $\mathbf{h}(\kk)$ and \ceq{eq:Df_h} for its relation to the quasiparticle Green's functions.
\bibitem{note9} Since there are, on average, three electrons per site,
there will be an equal number of sites with two and four electrons, corresponding to half of the total number of sites.
\end{thebibliography}
\end{document}